%% file: deuterium.tex
\documentclass[fleqn,usenatbib]{mnras}

\usepackage{newtxtext,newtxmath}

\usepackage[T1]{fontenc}

\DeclareRobustCommand{\VAN}[3]{#2}
\let\VANthebibliography\thebibliography
\def\thebibliography{\DeclareRobustCommand{\VAN}[3]{##3}\VANthebibliography}

\usepackage{graphicx}	%
\usepackage{amsmath}	%
\usepackage[separate-uncertainty=true]{siunitx}
\usepackage{booktabs}
\usepackage{longtable}

\makeatletter
\DeclareRobustCommand{\ion}[2]{%
  \mbox{#1\check@mathfonts\fontsize\sf@size\z@\selectfont #2}%
}
\makeatother

\DeclareRobustCommand{\kms}[1]{\SI{#1}{\km\per\second}}

\title[D/H towards PKS1937-101 with ESPRESSO]{Fundamental physics with ESPRESSO: a new determination of the D/H ratio towards PKS1937-101}

\author[F. Guarneri et al.]{
\parbox[t]{\textwidth}{
Francesco Guarneri,$^{1,2,3}$\thanks{E-mail: francesco.guarneri@inaf.it
\\ Based on observations collected at the European Southern Observatory, Chile, under the programme 0103.A-0512(A).}
Luca Pasquini$^{1}$,
Valentina D'Odorico$^{2,4,5}$,
Stefano Cristiani$^{2,4,6}$,
Guido Cupani$^{2,4}$,
Paolo Di Marcantonio$^2$, %
J. I. Gonz\'alez Hern\'andez$^{7,8}$, %
C. J. A. P. Martins$^{9,10}$, %
Alejandro Suárez Mascareño$^{7,8}$,
Dinko Milaković$^{4,2,6}$,
Paolo Molaro$^{2}$,
Michael T. Murphy$^{11,4}$, %
Nelson J. Nunes$^{12}$,
Enric Palle$^{7,8}$, %
Francesco Pepe$^{13}$,
Rafael Rebolo$^{7,8}$,
Nuno C. Santos $^{10,14}$,
Ricardo Génova Santos$^{7,8}$,
Tobias M. Schmidt$^{13}$, %
Sérgio G. Sousa$^{10}$, %
Alessandro Sozzetti$^{15}$, %
Andrea Trost$^{2,3}$
}
\\
\vspace*{6pt}\\
$^{1}$ ESO--European Southern Observatory, Karl-Schwarzschild-Strasse 2, 85748 Garching bei München, Germany \\
$^{2}$ INAF--Osservatorio Astronomico di Trieste, Via G.B. Tiepolo, 11, I-34143 Trieste, Italy \\
$^{3}$ Dipartimento di Fisica, Sezione di Astronomia, Università di Trieste, via G.B. Tiepolo 11, I-34131, Trieste, Italy \\
$^{4}$ IFPU--Institute for Fundamental Physics of the Universe, via Beirut 2, I-34151 Trieste, Italy \\
$^{5}$ Scuola Normale Superiore, P.zza dei Cavalieri, I-56126 Pisa, Italy\\
$^{6}$ INFN--National Institute for Nuclear Physics, via Valerio 2, I-34127 Trieste, Italy \\
$^{7}$ Instituto de Astrof{\'\i}sica de Canarias, E-38205 La Laguna, Tenerife, Spain \\
$^{8}$ Universidad de La Laguna, Dept. Astrof{\'\i}sica, E-38206 La Laguna, Tenerife, Spain \\
$^{9}$ Centro de Astrof\'isica da Universidade do Porto, Rua das Estrelas, 4150-762 Porto, Portugal \\
$^{10}$ Instituto de Astrofísica e Ciências do Espaço, CAUP, Universidade do Porto, Rua das Estrelas, 4150-762 Porto, Portugal \\
$^{11}$ Centre for Astrophysics and Supercomputing, Swinburne University of Technology, Hawthorn, Victoria 3122, Australia \\
$^{12}$ Instituto de Astrof\'isica e Ciências do Espa\c{c}o, Faculdade de Ci\^encias da Universidade de Lisboa,
Campo Grande, PT1749-016 Lisboa, Portugal \\
$^{13}$ Département d’astronomie, Université de Genève, Chemin Pegasi 51, CH-1290 Versoix, Switzerland \\
$^{14}$ Departamento de Física e Astronomia, Faculdade de Ciências, Universidade do Porto, Rua Campo Alegre, 4169-007 Porto, Portugal \\
$^{15}$ INAF - Osservatorio Astrofisico di Torino, via Osservatorio 20, 10025 Pino Torinese, Italy \\
}

\date{Accepted XXX. Received YYY; in original form ZZZ}

\pubyear{2024}

\begin{document}
\label{firstpage}
\pagerange{\pageref{firstpage}--\pageref{lastpage}}
\maketitle

\defcitealias{RS_17}{RS17}

\begin{abstract}
Primordial abundances of light elements are sensitive to the physics of the early Universe and can directly constrain cosmological quantities, such as the baryon-to-photon ratio $\eta_{10}$, the baryon density and the number of neutrino families.
Deuterium is especially suited for these studies: its primordial abundance is sensitive and monotonically dependent on $\eta_{10}$, allowing an independent measurement of the cosmic baryon density that can be compared, for instance, against the Planck satellite data. The primordial deuterium abundance can be measured in high \ion{H}{I} column density absorption systems towards distant quasars.
We report here a new measurement, based on high-resolution ESPRESSO data, of the primordial \ion{D}{I} abundance of a system at redshift $z \sim 3.572$, towards PKS1937-101. Using only ESPRESSO data, we find a \ion{D}{}/\ion{H}{} ratio of $\num{2.638\pm0.128e-5}$, while including the available UVES data improves the precision, leading to a ratio of $\num{2.608 \pm .102e-5}$. The results of this analysis agree with those of the most precise existing measurements. We find that the relatively low column density of this system ($\log{N_{\rm H_I}/ {\rm cm}^{-2}}\sim18 $) introduces modelling uncertainties, which become the main contributor to the error budget.
\end{abstract}


\section{Introduction}
The current standard cosmological model is remarkably able to describe our Universe from a few seconds after the Big Bang to the present day. Despite this, persistent issues have yet to be solved: we do not understand the nature of Dark Energy and Dark Matter, and tensions in the determination of key cosmological parameters have emerged \citep[see e.g.,][]{Abdalla_tensionH0}.

Investigating the early Universe can offer additional clues to the puzzle. A powerful probe of the physics of the early Universe is the Big Bang Nucleosynthesis (BBN), which predicts the abundances of the first light elements \citep[\ion{H}{}, \ion{D}{}, $^3$\ion{He}{}, $^4$\ion{He}{}, $^7$\ion{Li}{}. For a general review on the Big Bang Nucleosynthesis, see e.g.,][chapter 24 (Fields, Molaro, Sarkar)]{BBN_1, BBN_3, BBN_2, BBN_4}. These abundances are sensitive to the physics of the early Universe and to the cosmological model parameters: the number of neutrino families, for instance, can be constrained by measuring the He abundance \citep[see, e.g.,][]{peimber_He_neutrinos}. Another primordial element accessible with current facilities is D. Its abundance is of particular interest because it depends monotonically on the baryon-to-photon ratio and, as a consequence, on the baryon density. The primordial D abundance is measured in absorption systems seen in the spectrum of a background source, most commonly quasars, and computed under the assumption that the ratio of the chemical elements is equal to the ratio of the column densities of the observed neutral H and D, D/H = $N_{\rm D_I}$/$N_{\rm H_I}$, where $N$ is the column density of the absorbing gas. As discussed in \citet[and references therein]{cooke_18}, processes that weaken this assumption include the astration of deuterium as gas is processed by star formation, the relative reionization states of hydrogen and deuterium, or the depletion of deuterium onto dust grains. The first two effects are expected to be one order of magnitude below the uncertainties of current measurements at low metallicities. For the third, the amount of dust is generally small for systems in which deuterium is measured \citep{BBN_4}. 

As discussed in \citet{cooke_14, cooke_18}, the best systems for measuring the primordial deuterium abundance are those with \ion{H}{I} column density near the threshold of a Damped Lyman-$\alpha$ system (DLA, $\log(\mathrm{N_{H_I}})$\footnote{To simplify the notation, throughout the paper we use $\log{(N)}$ in place of $\log_{10}{(N/{\rm cm^{-2}}})$}\,$ \sim 20.3$). In this case, the Lorenzian damped wings allow a precise measurement of the total $\log{\rm N_{H_I}}$, while unsaturated \ion{D}{I} lines provide a robust measurement of $\log{\rm N_{D_I}}$.  
Unfortunately, DLA systems showing clear \ion{D}{I} absorption are rare: to date, only seven are known. Systems with column densities $\log{\rm N_{H_I}}\sim 17.5-19$ are more common \citep[][hereafter RS17]{RS_17}, but larger uncertainties are associated with the determination of the total \ion{H}{I} column density. A limited sample of DLAs at $z\sim2.5-3.0$ have been analysed since 1996, and the Precision Sample presented by \citet{cooke_18} provides a determination of the Deuterium abundance at the 1\% level.

In the following, we report of a new measurement of \ion{D}{I}/\ion{H}{I} in the absorption system at redshift $z = 3.572$ toward PKS1937-101 (J2000 19:39:57.26 -10:02:41.5) based on new data with higher signal-to-noise ratio (S/N) and resolution compared to previous data sets. This system is peculiar due to the low metallicity ($Z/Z_\odot\lesssim-2$), relatively low column density, and high redshift. 

Prior to the new observations presented here, PKS1937-101 was observed using HIRES \citep[][]{HIRES} and UVES \citep[][]{UVES} spectrographs. These observations returned a value of \ion{D}{I}/\ion{H}{I} = \num{2.62 \pm 0.051e-5}  \citep{RS_17}. However, both instruments have been shown to suffer from wavelength calibration issues \citep[][]{spectrographs_wave_distortions}, introducing long-range wavelength scale distortions of $\pm 200\ {\rm m\ s^{-1}}$ into the quasar spectra, and also covered a relatively short wavelength range ($\sim$400-660 nm for UVES and $\sim$380-680 nm for HIRES). To overcome these shortcomings, a new spectrum of PKS 1937-101 was collected in 2019 using the Echelle SPectrograph for Rocky Exoplanet and Stable Spectroscopic Observations \citep[ESPRESSO,][]{ESPRESSO}. ESPRESSO's wavelength calibration is significantly better than those of UVES and HIRES \citep{Schmidt_2021}, and its extended wavelength coverage (380-780 nm) helps identifying interlopers and systems that could contaminate the \ion{D}{I} and \ion{H}{I} lines not identified previously. In addition, the higher resolution of ESPRESSO allows to better resolve the velocity structure of the system.

The paper is organised as follows: in Sect. 2 we describe the data collection and reduction; in Sect. 3 we describe the analysis of the data, the resulting \ion{D}{I}/\ion{H}{I} ratio, and uncertainties in the modelling of the data. In Sect. 4, archival datasets are included in the analyses while in Sect. 5 we describe a Cloudy model of the system. In Sect. 6, we compare our new result to other measurements, and present our conclusions in Sect. 7.

\section{Observations and data reduction}
PKS1937-101 is a bright (V$_\mathrm{mag}$ = 16.7) quasar at redshift $z=3.787$. It was observed on 22 August 2019 as a part of the ESPRESSO consortium's guaranteed time observation programme for a total of 17800 seconds spread over 5 exposures (see Table~\ref{tab:obsSummary}). The exposures were taken using the 4-UT mode of ESPRESSO, in which all four Unit Telescopes of the VLT simultaneously observe the target and their light is combined at the ESPRESSO front end. This configuration effectively makes VLT equivalent to a 16m telescope in area and was chosen to increase the final S/N in the data. Pixel binning along the spatial direction was 8 pixels and binning along the spectral direction was 4 pixels (the so-called `multi-MR84' mode), providing a nominal spectral resolution of $\sim70000$ and a wavelength coverage from $380$ nm to $780$ nm.

\begin{table}
    \centering
    \begin{tabular}{cccc}
    \toprule
    Observing time & Exp. time & S/N & S/N$_{{\rm Ly}\alpha}$ \\
    (UTC) & (s) & (pix$^{-1}$) & (pix$^{-1}$) \\
    \midrule
    2019-08-21T23:30:03.725 & 3600 & 20 & 26 \\
    2019-08-22T00:41:08.617 & 4100 & 25 & 34 \\
    2019-08-22T01:50:11.649 & 4100 & 28 & 38 \\
    2019-08-22T02:59:15.783 & 3600 & 28 & 37 \\
    2019-08-22T04:13:54.673 & 2400 & 22 & 27 \\
    \bottomrule
    \end{tabular}
    \caption{The final spectrum used for the analysis consists of five combined exposures. Here, we summarise relevant information. The S/N column is the median signal-to-noise ratio along the spectrum, while the S/N$_{{\rm Ly}\alpha}$ is the median S/N in the continuum just redward of the Ly$\alpha$ absorption system analysed in this work, at $\lambda \sim 5562$ \AA.}
    \label{tab:obsSummary}
\end{table}

Data were reduced using the standard ESO ESPRESSO data reduction software (DRS) version 3.0.0. We summarise the relevant steps and refer the interested reader to \citet{ESPRESSO} for more details. The main steps performed are: i) bias, dark and inter-order background subtraction; ii) optimal extraction, using a modified version of the \citet{Zechmeister_optimalExtraction} algorithm; iii) creation of extracted spectra, with associated error and quality maps; iv) flat fielding and de-blazing; v) wavelength calibration, using either Laser Frequency Comb (LFC, calibration chosen for this work) or Thorium-Argon lamps combined with Fabry-P\'erot sources; vi) extraction of the sky spectrum, and creation of a sky-subtracted 2D spectrum. The sky signal was subtracted using the smoothing recipe of the ESPRESSO DRS pipeline. This mode performs first a sliding average of the sky spectrum and then subtracts if from the science spectrum.

\begin{figure*}
    \centering
    \includegraphics[width=\textwidth]{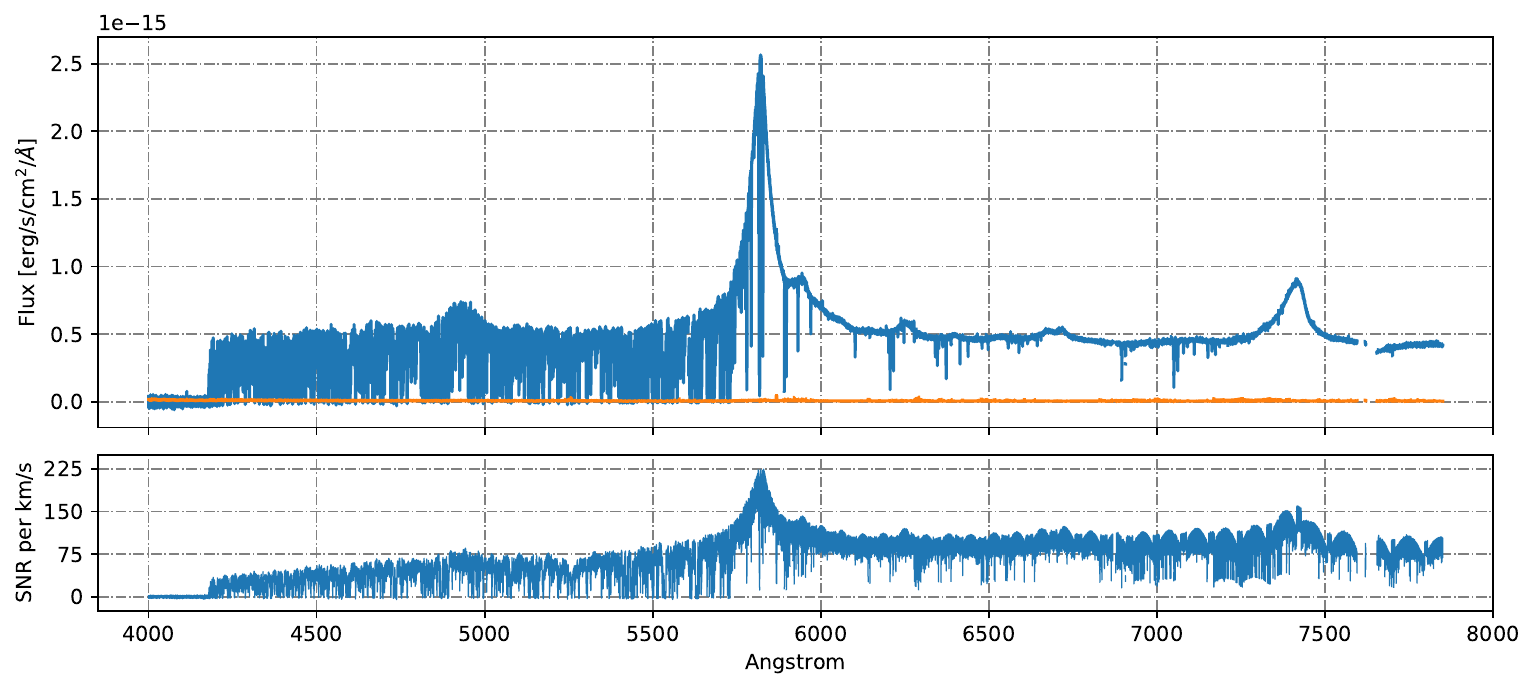}
    \caption{Top panel: wavelength calibrated ESPRESSO spectrum of PKS1937-101. The blue line represents the flux, while the orange line represents the associated error. Bottom panel: formal S/N per \kms{} along the spectrum.}
    \label{fig:ShowSpec}
\end{figure*}

The five wavelength-calibrated frames were combined using the python software package \textsc{Astrocook} \citep{Astrocook}, resulting in a spectrum with a S/N per km s$^{-1}$ of around 100 redwards of the Lyman-$\alpha$ (Ly$\alpha$) emission line, and from $\sim30$ to $\sim90$ in the Ly$\alpha$ forest (Fig. \ref{fig:ShowSpec}, bottom panel). The spectrum was rebinned to a wavelength grid constant in velocity space, with pixel size corresponding to \kms{2.0}, providing a 2-pixel sampling of the full-width at half-maximum (FWHM) of the resolution element. The formal flux error vector, derived following standard propagation of the uncertainties of the five combined frames, was found to be $\sim 1.3\times$ lower than what would be expected from the observed fluctuations in flux over wavelength regions free from any features. This discrepancy does not seem related with the procedure of combination of the single spectra, but could be due to the treatment of the dark current in the reduction pipeline. A more detailed investigation of this problem goes beyond the scope of this work, as a consequence we decided to replace the error vector by the reduced root mean square error of pixel fluxes:
\begin{equation}
    RMSE_i = \sqrt{\frac{1}{\sum_j w_j} \frac{1}{\sum_j p_j - 1} \sum_j^n w_j (f_j - F_i)^2}.
\end{equation}
Above, $F_i$ is the flux of an $i$\textsuperscript{th} pixel in the combined spectrum, $f$ are the fluxes of $n$ pixels that contribute to it, $w$ the weights assigned to each of the $n$ pixels and $p$ the fraction each of the $n$ pixels contributing to $F_i$.

The quasar continuum was estimated using \textsc{Astrocook} by applying an iterative sigma-clipping procedure to remove absorption features while masking emission lines. The continuum was then computed as the average, unclipped flux, subsequently smoothed in velocity space with a Gaussian filter (FWHM = 200 \kms{}). This produces reliable results redward of the Ly$\alpha$ emission, where absorption lines are generally narrow and sparse. On the other hand, on the blue side of the Ly$\alpha$ emission line, the recipe systematically underestimates the continuum. We thus manually corrected the continuum level where necessary. While this is a subjective process, it is fully repeatable owing to the capabilities of \textsc{Astrocook}. We briefly discuss the effect of continuum placement on the line fit in Sect. \ref{sec:ContinuumPlacement}.  

\section{Data analysis}
\label{ref:mainModel}
We fit the absorption lines of the system at $z \sim 3.572$ using VPFIT 12.4 \citep{VPFIT}. VPFIT is a software package used for Voigt profile fitting of absorption spectra. Each Voigt profile is parameterised by an atomic species, with corresponding atomic parameters, and up to four free parameters: the position in redshift space, the column density of the absorber, and the broadening, or Doppler, parameter. Atomic parameters (such as the laboratory wavelength and oscillator strength) are provided within the VPFIT package and are a compilation of several sources in the literature. The free parameters are the redshift of the absorber or $z$, its column density or $\log(N)$, and up to two parameters describing the line broadening \textit{b}. Two terms can in principle contribute to the Doppler parameter, one accounting for the turbulent ($b_\mathrm{turb}$) and one for the thermal broadening ($b_\mathrm{term}$). Should the profile be broadened by fully turbulent or fully thermal broadening mechanisms, three parameters are needed to fully describe a Voigt profile. The presence of multiple species with different atomic masses in the same absorption system, however, allows for disentangling the relative contributions of the two broadening mechanisms, according to the relation:
\begin{equation}
    b_\mathrm{tot}^2 = b_\mathrm{turb}^2 + b_\mathrm{ther}^2 = b_\mathrm{turb}^2 + \frac{2kT}{m}
\end{equation}
where $k$ is the Boltzmann constant, $T$ is the temperature of the gas and $m$ is the mass of the atomic species of interest. 

Provided with a user-supplied model, VPFIT optimises the free parameters using nonlinear least squares minimisation. It then reports the best-fit value for the free parameters, the corresponding errors, the $\chi^2$ and the reduced $\chi_\nu^2$ (i.e., the $\chi^2$ divided by the number of degrees of freedom, $\nu$). $\chi^2_\nu$ was used to guide the model building process, that is a component was added only when it was deemed necessary to reach $\chi_\nu^2\approx1$.

VPFIT also requires information about the instrumental profile, which is convolved with the theoretical models before they are compared to the observed spectrum during parameter optimisation. Here, the instrumental profile of ESPRESSO is assumed to be a Gaussian with a FWHM of \kms{4.28}, corresponding to the nominal resolution of the adopted ESPRESSO configuration. While the instrumental profile of ESPRESSO is known to depart from a pure Gaussian shape \citep{Schmidt_2021,2022arXiv221202458M}, we retain the Gaussian form because we lack information on its exact shape, but acknowledge that this might leave certain systematic effects in our analysis. For example, if ESPRESSO's instrumental profile varies across the detector in a non-trivial way, as was recently found for a high resolution spectrograph with a similar optical design to ESPRESSO \citep{2023arXiv231105240M}, there may be intra-order and inter-order wavelength distortions that were left unaccounted for in this analysis.

Due to the large width and saturation of the \ion{H}{I} line of the studied system, it is difficult to accurately identify its velocity structure and the number of components needed to carry out a proper fit. On the other hand, metal lines are easier to model for this system: they are unsaturated and narrower than the Ly$\alpha$ profiles. We derived a model for the saturated \ion{H}{I} regions assuming that low-ionisation metal species and hydrogen share the same velocity structure. In practice, we first modelled the metal lines to obtain the velocity structure of the system and the first guesses for the redshift and Doppler parameter of each \ion{H}{I} component. Next, we fitted both metals and hydrogen transitions simultaneously, tying the redshift and Doppler parameters of metals, deuterium and hydrogen together. The redshift parameters were required to be the same for all species, but otherwise free to vary during the fitting procedure. The same applies to the turbulent Doppler parameter and the temperature for each species.

\subsection{Metal absorption lines}
Starting from the metal lines detected in the studied system by \citetalias{RS_17}, we searched the ESPRESSO spectrum for all associated ionic transitions. Thanks to the extended wavelength range of the ESPRESSO spectrum, we were also able to detect the \ion{Si}{II} 1526 \AA{} and the \ion{C}{IV} doublet at 1548, 1550 \AA. 
We considered for our model 6 low-ionisation transitions: \ion{C}{II} 1334 \AA{}, \ion{Si}{II} 1193 \AA{}, \ion{Si}{II} 1260 \AA{}, \ion{Si}{II} 1304 \AA{}, \ion{Si}{II} 1526 \AA{} and \ion{Fe}{III} 1122 \AA. 

Additional metal absorption lines (namely, \ion{C}{III} 977 \AA{}, \ion{Si}{III} 1206 \AA{}, \ion{C}{IV} 1548, 1550 \AA{} and \ion{Si}{IV} 1393, 1402 \AA) are visible in the spectrum, but were not included in the model. \ion{C}{III} and \ion{Si}{III} are in the Lyman forest, where, for both lines, interlopers affect the velocity profile. High-ionisation metals, such as \ion{C}{IV} and \ion{Si}{IV}, may not share the same velocity structure as low-ionisation species. However, we checked and confirmed that the components found for the other metals were consistent with the \ion{C}{III}, \ion{C}{IV} and \ion{Si}{III}, \ion{Si}{IV} absorption.

Finally, some commonly found metal transitions, such as \ion{O}{I} 1302 \AA, are not clearly visible in the spectrum or are heavily contaminated by telluric lines (e.g., \ion{Al}{II} 1670 \AA). Forcing the presence of \ion{O}{I} 1302 \AA, while fitting for other metal lines, returns a nonzero summed column density equal to $\log{N_{\rm OI}}$ = \num{12.1\pm0.1}. This value is consistent with the previous determination of \citetalias{RS_17}, obtained under slightly different hypothesis, suggesting that the column density determination for \ion{O}{I} 1302 \AA{} is a reliable upper limit (we consider this value as an upper limit as these lines are not detected in the spectrum and are force-fitted using VPFIT based on knowledge of the position of the other metal transitions).

\begin{figure}
    \centering
    \includegraphics[width=\columnwidth]{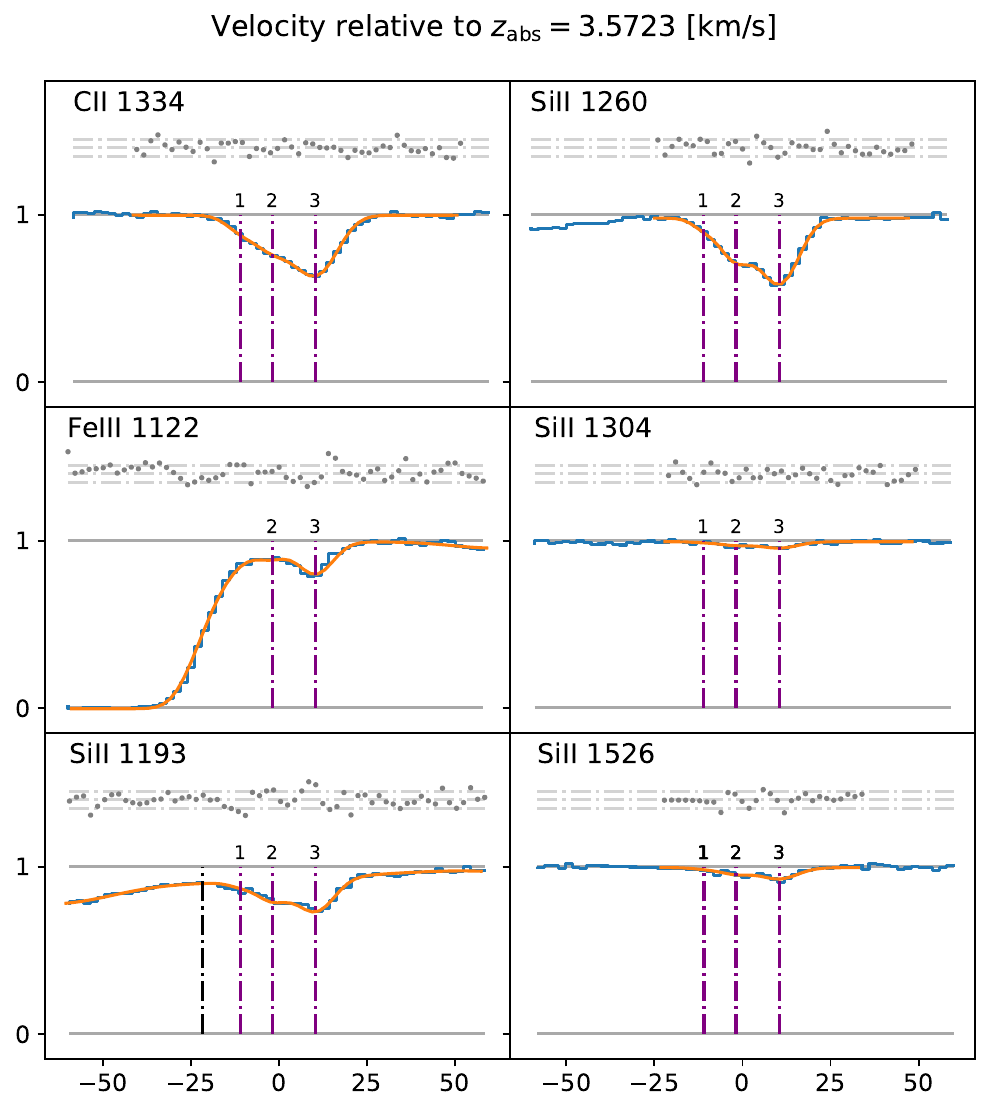}
    \caption{Low ionisation metal lines used in the model to constrain the velocity structure of the hydrogen at the same redshift. Absorption components are marked with purple vertical lines, and the three components comprising the model are visible in, e.g., the \ion{C}{II} or \ion{Si}{II} 1260 \AA{} lines. Black lines mark the position of unrelated \ion{H}{I} absorption systems, while the grey line are the normalised residuals (data-model)/error, and the horizontal grey dot-dashed line are the 1-$\sigma$ limits.}
    \label{fig:lowIonizationMetalLines}
\end{figure}

We initially only consider metal lines and produce a model of the six listed transitions. Three velocity components, marked 1, 2, 3 in Fig. \ref{fig:lowIonizationMetalLines}, are needed.

\subsection{Determination of the Deuterium abundance}
\label{sect:DoH_ESPRESSO}
We applied the model derived from metal species to \ion{H}{I} and \ion{D}{I}. An additional line is visible blueward of the main \ion{H}{I} absorption in the Ly$\alpha$, Ly$\beta$, and Ly$\gamma$ lines (Fig. \ref{fig:ModelEspressoOnly}). There are strong indications that this line arises from \ion{D}{I} absorption: i) it is significantly broader than other metal lines seen in the spectrum but narrower than the associated Lyman series absorption; ii) the shift between this line and the main \ion{H}{I} absorption is $\sim$\kms{-82}, consistent with the expected shift between \ion{D}{I} and \ion{H}{I}.

We assumed that low-ionisation metals, hydrogen and deuterium share the same number of components, redshift and Doppler broadening. Column densities were instead free to vary independently. We fit low ionisation metal lines together with the Lyman series (Ly$\alpha$ to Ly9) and the corresponding deuterium series. In order to achieve a more accurate estimate of the total column density of \ion{H}{I} and \ion{D}{I}, we solved for their respective summed column densities across the absorption system (i.e., we directly found the summed column density for all components of the model). We forced all sub-components to have the same \ion{D}{I}/\ion{H}{I} ratio \citep[an hypothesis commonly assumed; see, e.g., ][]{cooke_18}: this implicitly assumes that \ion{D}{I} depletion, if present, affects equally all subcomponents.

Several regions require additional unrelated \ion{H}{I} absorbers to obtain a good fit. When possible, these are modelled as Ly$\beta$ or Ly$\gamma$ lines with an associated Ly$\alpha$ transition at a longer wavelength. This provides more robust constraints on the number, column density, and position of these lines compared to the use of only Ly$\alpha$.

\begin{figure*}
    \centering
    \includegraphics[width=\textwidth]{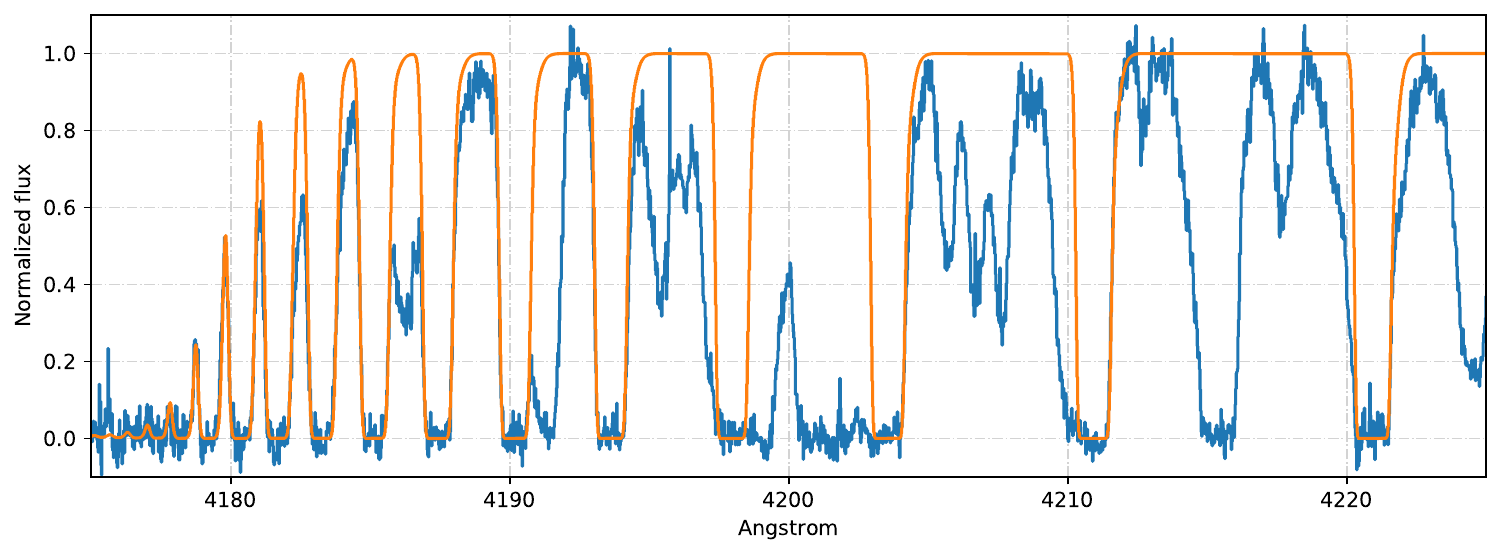}
    \caption{ESPRESSO model (orange line) over-plotted onto the high order \ion{H}{I} Lyman series lines, from Lyman 10 to Lyman 20 (blue line). The model appears to be consistent with the data.
    }
    \label{fig:HighOrderLines}
\end{figure*}

\begin{figure*}
    \centering
    \includegraphics[width=\textwidth]{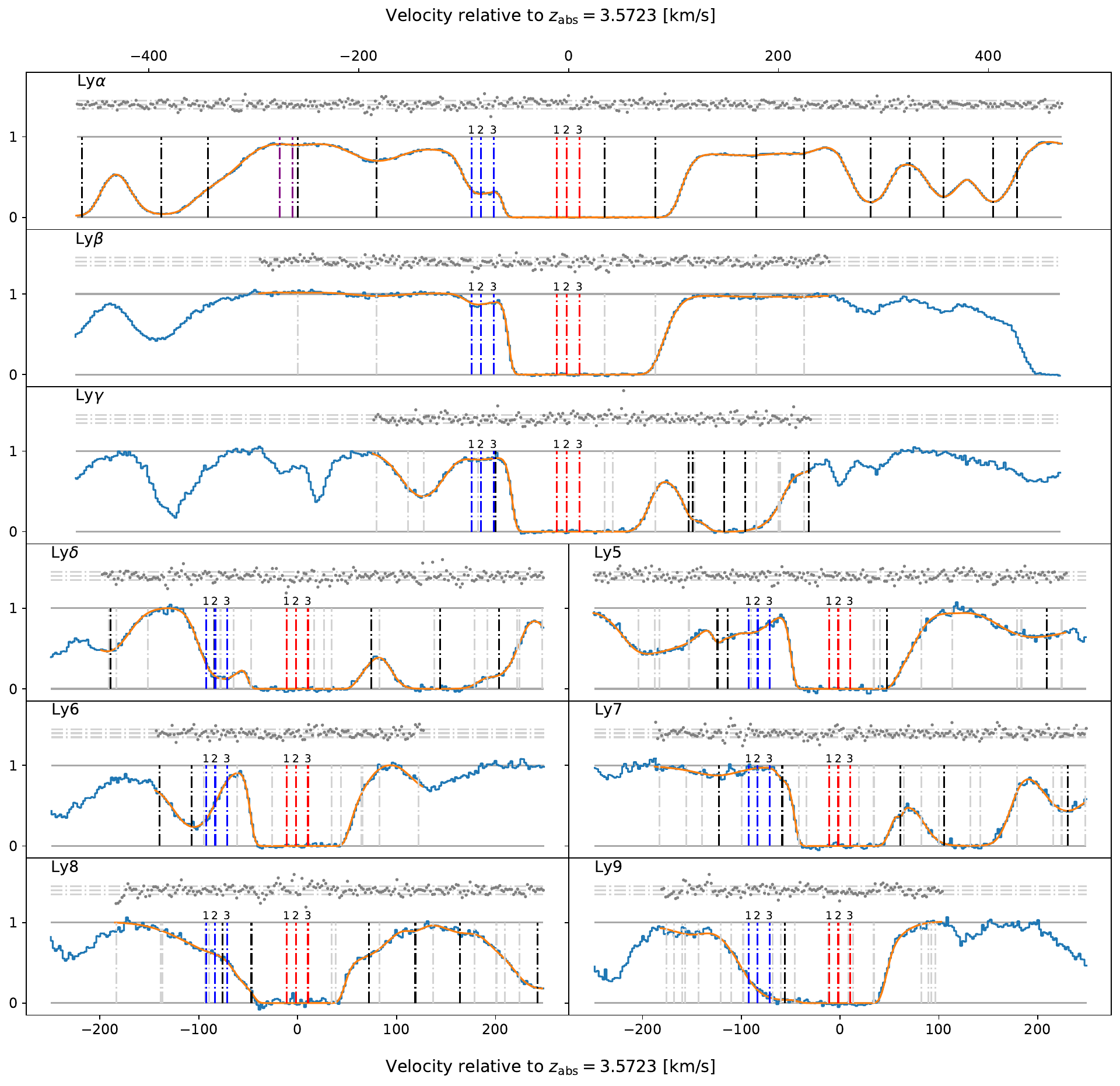}
    \caption{Model fitted to the ESPRESSO data only. The blue line shows the spectrum of PKS1937-101, while the orange line shows the model. Residuals are shown above each region, over-plotting the $\pm1\sigma$ limits.
    Black dash-dotted lines mark the position of interlopers (i.e., an unrelated \ion{H}{I} system along the same line of sight). We only show interlopers added in each region: the remaining absorption is due to Lyman lines at higher redshift. Purple lines in the Ly$\alpha$ region mark the position of the \ion{Fe}{III} at the same redshift of the main system. The red and blue lines mark the positions of the \ion{H}{I} and \ion{D}{I} components, respectively. Each component of the model is marked with a number (see text).}
    \label{fig:ModelEspressoOnly}
\end{figure*}

We have not explicitly fit transitions higher than the Lyman 9, albeit visible in the spectrum, because the placement of the continuum is uncertain and several interlopers make it difficult to separate the contribution from the main system and unrelated ones. However, we checked that our model does not produce obvious defects when superimposed on those transitions.
In Fig. \ref{fig:HighOrderLines}, we plot the model and the ESPRESSO spectrum for the Lyman series from Lyman 10 to 20. The two are visually consistent with each other.

Finally, we fit the model to the data and determine the column densities of \ion{H}{I} and \ion{D}{I} finding $\log(\ion{N}{\ion{H}{I}}) = \num{17.923\pm 0.015}$, $\log(\ion{N}{\ion{D}{I}}) = \num{13.345 \pm 0.015}$ and a \ion{D}{I}/\ion{H}{I} ratio of $\num{2.638 \pm 0.128e-5}$. The temperature of the gas, averaged across the three components, is \num[]{1.68e4} K. The final $\chi_\nu^2$ for the fit is 0.96. We report in Tab. \ref{tab:modelTransition} and \ref{tab:UnrelatedInterlopers} all relevant parameters for all components.

\subsection{Modelling uncertainty}
\label{sect:model_unc}
We now explore various systematic effects that may have affected our measurements.

\subsubsection{Higher excitation metal lines}
Additional lines are available in the spectrum, besides those considered in the modelling: the \ion{C}{III} 977, \ion{Si}{III} 1206, \ion{Si}{IV} and \ion{C}{IV} doublets. We check that our model is consistent with these absorption systems, finding that extra components are needed to model these lines.

\begin{figure}
    \centering
    \includegraphics[width=\columnwidth]{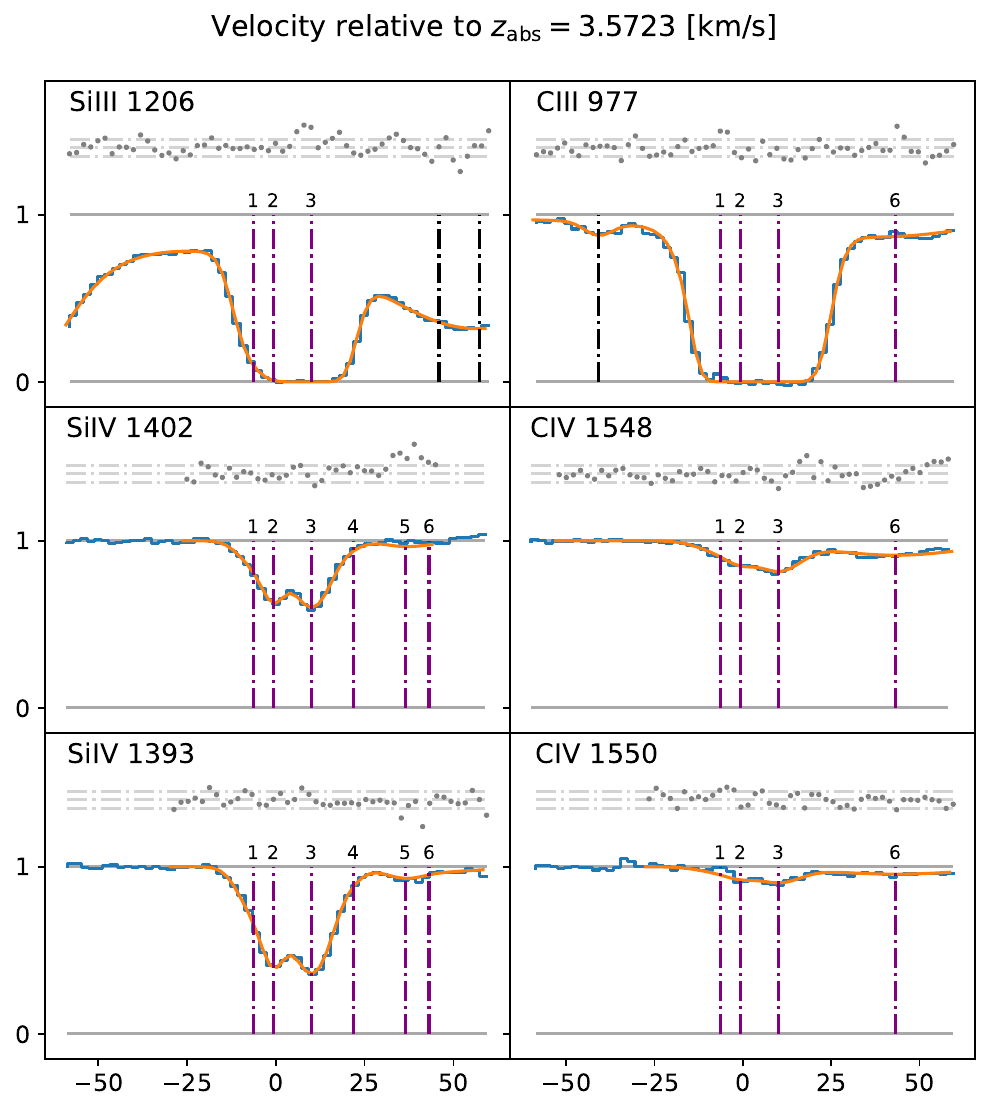}
    \caption{Additional metal lines of the system at $z=3.5723$, not included in the model due to contamination or differences in ionisation potential. As for the other figures, purple lines mark the position of metal lines, numbered on top, while black lines the position of possibly unrelated absorbers, modelled as Ly$\alpha$.}
    \label{fig:CIV_SiIV}
\end{figure}

\begin{itemize}
    \item \ion{Si}{IV}, \ion{C}{IV}: \citetalias{RS_17} include SiIV in the fitting model for \ion{H}{I} and \ion{D}{I}, despite the differences in the ionisation potential. Following their approach, we check if the three-component model is adequate to describe the \ion{Si}{IV} and the \ion{C}{IV} doublets.
    
    We found that two and three more components were required to describe the absorption lines. One of these, marked 4 in Fig. \ref{fig:CIV_SiIV}, is similar to the fourth component found by \citetalias{RS_17}. The components marked as 5 and 6, on the other hand, were previously not identified as part of the absorption system due to lack of wavelength coverage in the UVES and HIRES spectra. These two components are not visible in any of the lower-ionisation metal lines at the same redshift, supporting the hypothesis that high-ionisation lines have a different velocity structure and the choice of not including them in the fit model for hydrogen and deuterium.
    
    \item \ion{C}{III} 977, \ion{Si}{III} 1206 \AA: The three components present in all metal transitions can account for the strong absorption trough seen in Fig. \ref{fig:CIV_SiIV}. However, both lines are saturated and fall in the Lyman forest, and thus are likely contaminated by \ion{H}{I} lines (black, dash-dotted lines in Fig. \ref{fig:CIV_SiIV}). 
    This is true especially for \ion{Si}{III}, which shows a strong absorption trough that is not evident in any other metal line. 
    Interpreting the velocity structure of \ion{C}{III} is harder: the absorption to the left and right of the strong trough seen in Fig. \ref{fig:CIV_SiIV} can either be associated to the fifth and sixth components discussed in the previous bullet point or to an additional component, marked by the number 7 in Fig. \ref{fig:CIV_SiIV}, to \ion{H}{I} interlopers, or to a mix of both.
\end{itemize}

\subsubsection{A model with four components}
\label{sect:modelFourComponents}
\citetalias{RS_17} present a model comprising four components (see, e.g., Fig. 1 of this reference). However, our model considers only three. We found that a fourth component is only required to adequately fit the \ion{Si}{IV} doublet. 
Taking into account only low-ionisation metal species, namely the \ion{Si}{II}, \ion{C}{II}, and \ion{Fe}{III} transitions, VPFIT rejects the fourth component entirely. On the other hand, including the \ion{H}{I} lines makes it so that the fourth component is kept, although with larger errors (by a factor of $\sim 10$) in column density, redshift, and Doppler parameters compared to the other three. In this case, the resulting \ion{D}{I}/\ion{H}{I} is \num{2.634 \pm 0.230e-5}, essentially unchanged with respect to the three-component model, but with significantly higher uncertainty.

If, in addition to including the fourth component, we also include all available metal species (\ion{C}{III} 977, \ion{Si}{III} 1206 and the \ion{C}{IV} and \ion{Si}{IV} doublets) and refit the model, we find a comparable hydrogen and deuterium column densities $\log(\ion{N}{\ion{H}{I}}) = \num{17.925\pm 0.015}$, $\log(\ion{N}{\ion{D}{I}}) = \num{13.343 \pm 0.016}$ with slightly larger uncertainties and lower \ion{D}{I}/\ion{H}{I} ratio of $\num{2.618 \pm 0.133e-5}$.

\subsubsection{Contamination of \ion{H}{I} lines in the Ly$\alpha$ region}
Contamination by hydrogen interlopers in regions with deuterium absorption is possible, although unlikely ($<3.7\%$, as discussed in \citetalias{RS_17}). Nonetheless, we investigate how the inclusion of an additional line in the \ion{D}{I} region affects the results. We proceed as follows: first, we assume that no deuterium is present and model the absorption trough blueward of the main Ly$\alpha$ system with a single \ion{H}{I} component. The parameters fitted for this component are: $\log ({\rm N_{H_I}}) = 13.34$, $b = \kms{18.09} $, and $z = 3.57102$. We then reintroduce the deuterium and refit the model including both the deuterium and the additional \ion{H}{I} line. We use as starting parameters for the additional \ion{H}{I} line the results of the aforementioned fit. Under the same assumption used for the main model, we find $\log(\ion{N}{\ion{H}{I}}) = \num{17.919\pm 0.022}$, $\log(\ion{N}{\ion{D}{I}}) = \num{13.341 \pm 0.037}$ and \ion{D}{I}/\ion{H}{I} $ = \num{2.644 \pm 0.264e-5}$.

\subsubsection{Choice of the fitting region}
\label{sec:fittingRegion}
An additional source of uncertainty is related to the choice of the fitting region, combined with the inclusion of continuum adjustments allowed by the VPFIT model. In particular, considering the Ly$\alpha$ region, we find that restricting the fitting region to $\pm400$ \kms{} (Fig. \ref{fig:ModelEspressoOnly}, upper panel) leads to a lower \ion{H}{I} column density, $\log({\rm N_{H_I}})=\num{17.905\pm0.023}$ and consequently to a higher \ion{D}{I}/\ion{H}{I} ratio $\num{2.76\pm0.16e-5}$. This is caused by the exclusion of regions where the unabsorbed continuum is higher than in the $\pm400$ \kms{} range: while fitting the model, VPFIT adjusts the continuum to be lower. Although this is possible, it is unlikely that this is the case, as it would require the continuum to change on small scales.

\subsubsection{Continuum placement} \label{sec:ContinuumPlacement}
\begin{figure}[htb]
    \centering
    \includegraphics[width=\columnwidth]{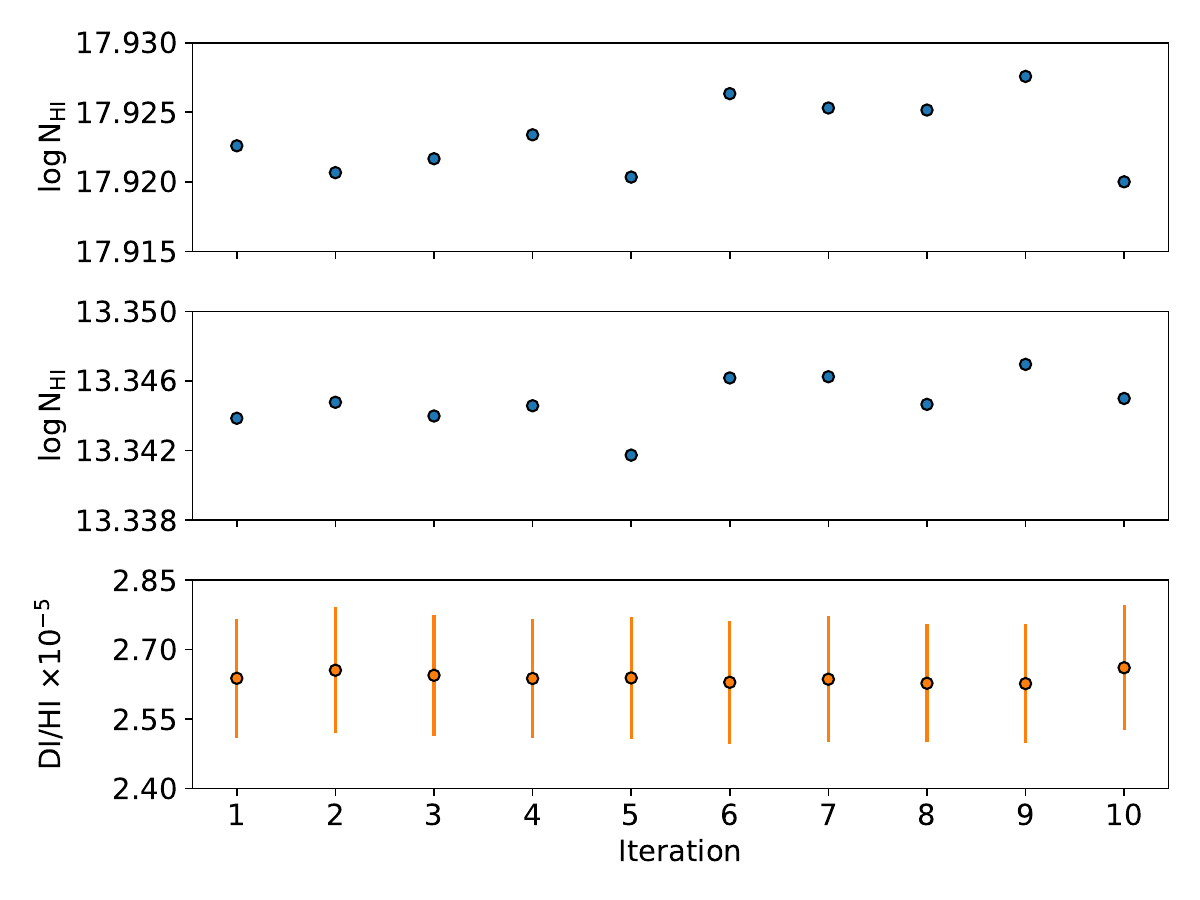}
    \caption{Top panel: \ion{H}{I} column density for each iteration. Middle panel: same, for deuterium column density. Lower panel: same, but for the resulting \ion{D}{I}/\ion{H}{I}.}
    \label{fig:iteration_res}
\end{figure}

In addition to allowing small continuum adjustments in the VPFIT model in each region, we investigate whether the position of the continuum influences the final determination and its uncertainty. To quantify this effect on the final \ion{D}{I}/\ion{H}{I} ratio, we randomly varied the placement of the continuum and re-fit the model on the newly normalised spectra. Ten different realisations were produced. 
To produce each realisation, we multiplied the continuum of the spectrum by a slowly varying function, generated by extracting points from a Gaussian distribution centred at unity and with a standard deviation $\sigma=0.05$. Each point was associated with a wavelength, using a range from 3000\AA{} to 8000\AA{} and a step of 50\AA. Finally, points were interpolated with a smoothing spline and evaluated on the same wavelength scale of ESPRESSO.
The model presented in the previous sections was finally fitted separately to each of the ten realisations. We computed the \ion{D}{I}/\ion{H}{I} ratio for each realisation: their average is $\num{2.639\pm0.010e-5}$, where the error is taken as the standard deviation of the ten iterations. The uncertainty associated with the continuum is one order of magnitude lower than the statistical uncertainty of the fit. Summing the contributions in quadrature increases the final uncertainty on the \ion{D}{I}/\ion{H}{I} ratio by $\sim 0.5 \%$. The mean error (that is, the average of the errors estimated from the ten iterations) is 0.131. Fig. \ref{fig:iteration_res} shows, per each iteration, the column density of deuterium and hydrogen, and the corresponding \ion{D}{i}/\ion{H}{I} ratio.

\section{Including previous datasets: UVES and HIRES}
PKS1937 had also been observed with the HIRES and UVES spectrographs. We have analysed these data trying to see if they could be useful for improving the Deuterium fit.

\subsection{Archival data}
PKS1937 has been observed in 2006 and 2007 with UVES \citep{UVES}, proposal 077.A-0166(A) (P.I.\ R.\ F.\ Carswell). During the execution of the programme, the object was observed with a resolution of ${\rm R} = 45000$, for a total of \num{5.4e4} seconds spread among 10 individual exposures of \num{5.4e3} seconds. The spectrum used in this work was taken from the SQUAD dataset \citep{squad_dr1}, a large, public collection of fully reduced, wavelength calibrated QSO spectra observed with UVES. The combined UVES spectrum has roughly half the signal to noise ratio per \kms{}, compared to the ESPRESSO spectrum, averaging $\sim37$ in the Lyman forest and $\sim65$ redward of the Ly$\alpha$ emission line. A detailed description of the steps taken to produce the catalogue, reduce each science frame, and produce a stacked spectrum is available in the reference paper.\footnote{Details of the data processing for all SQUAD QSOs are also available online at the SQUAD web page (\href{https://github.com/MTMurphy77/UVES_SQUAD_DR1}{github.com/MTMurphy77/UVES\_SQUAD\_DR1}).}

Science frames were reduced using the ESO Common Pipeline Library and combined using UVES\_popler \citep{uves_popler}. The same tool was also used to fit a continuum on the combined spectrum. To do so, data is broken into overlapping chunks of a user-defined width, and a polynomial is fitted on each chunk. To form a smooth continuum, continua from adjacent chunks are averaged together. It should be noted that the automatic procedure does not work well in the Lyman forest, due to the small amount of un-absorbed pixels. For quasars in SQUAD, the continuum in the region of the Lyman forest was manually refitted, selecting seemingly unabsorbed peaks in the Lyman forest and interpolating between them with a low-order polynomial. As for \textsc{Astrocook}, the procedure is subjective but fully repeatable, thanks to UVES\_popler log files.

In addition to UVES, PKS1937-101 has been extensively observed using HIRES, mounted on the Keck Telescope in Hawaii. Reduced, wavelength-calibrated, and continuum normalised archival data are available as part of the KODIAQ dataset \citep{kodiaq_dr3}, and were retrieved from SpecDB \citep{specdb, igmspec}.
Three co-added spectra are available, resulting from observational programmes carried out between 1997 and 2005 (P.I.\ L.\ Cowie, N.\ Crighton, and D.\ Tytler). We refer the interested reader to the KODIAQ paper \citep{kodiaq_dr3} for details about data reduction and processing.

\begin{figure}
    \centering
    \includegraphics[width = \columnwidth]{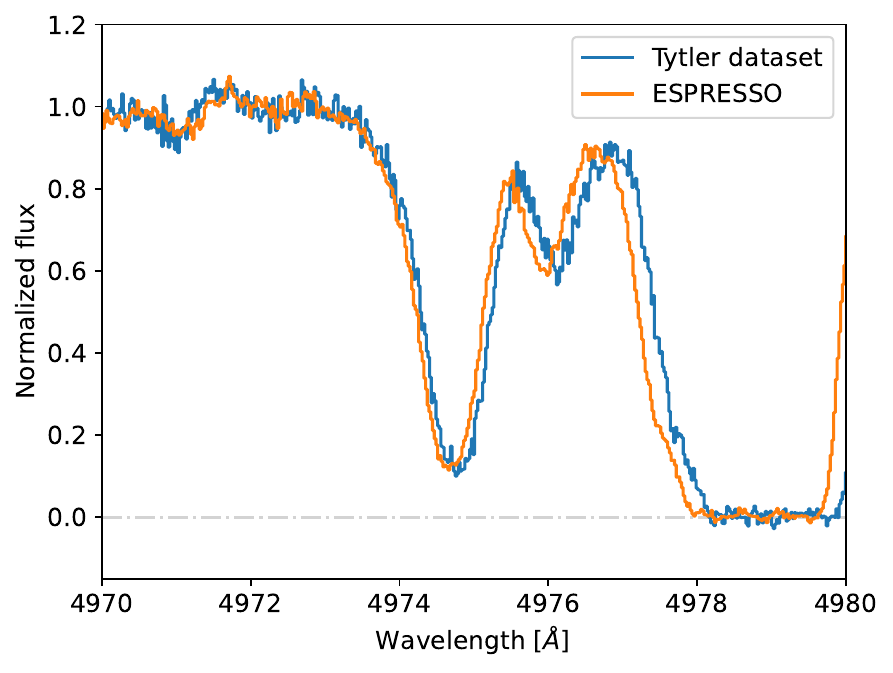}
    \caption{Example of wavelength calibration issues between the HIRES (blue) and ESPRESSO datasets (orange).}
    \label{fig:TytlerWaveProb}
\end{figure}

Of the three datasets, wavelength calibration issues in one of the combined spectra were noted (P.I.\ D.\ Tytler, Fig. \ref{fig:TytlerWaveProb}) in the form of a wavelength-dependent shift between the ESPRESSO and the HIRES data. Thus we discarded it and limited our analysis to the remaining two. The signal to noise ratio per \kms{} of the two spectra combined is on average $\sim$ 40 in the forest, 60 red-ward of the Ly$\alpha$ emission line. Finally, it should be noted that the algorithm used to estimate the continuum on HIRES data is again different from the one used for UVES or ESPRESSO spectra. In the case of HIRES, data are continuum normalised on an order-by-order basis by fitting Legendre polynomials (with orders varying from 4th to 12th) in regions deemed free of absorption. Employing three different algorithms might produce different artefacts or systematics in the combined fit.

\subsection{Fitting the ESPRESSO model on the UVES data}
\label{sect:modelOnUVES}
Before fitting the model on UVES data, we took advantage of the significantly better ESPRESSO wavelength calibration to correct the wavelength grid of the UVES data, which is known to be problematic \citep[e.g.,][]{RamahaniUvesShift}. We did this by computing velocity shifts between UVES and ESPRESSO data, with ESPRESSO being chosen as the `correct' one, and shifting the UVES wavelength scale to match that of ESPRESSO. When this is done, the UVES and ESPRESSO spectra can be fitted together without additional parameters of velocity shifts between them. 

The shifts were calculated by cross-correlating the two spectra in the velocity space, which were rebinned to a common velocity grid, with a pixel size of 1.4 km s$^{-1}$. The choice of the pixel size did not affect the resulting shifts, as long as it is not too large (> \kms{4.5}). The two spectra were then subdivided into chunks, and each pair of corresponding chunks was cross-correlated with each other. We computed the cross-correlation considering relative shifts between -5 and 5 km s$^{-1}$, with a fixed step of $10^{-3}$ km s$^{-1}$. The velocity shift corresponding to the maximum of the cross-correlation function was taken as the velocity shift between two chunks.

\begin{figure}
    \centering
    \includegraphics[width=\columnwidth]{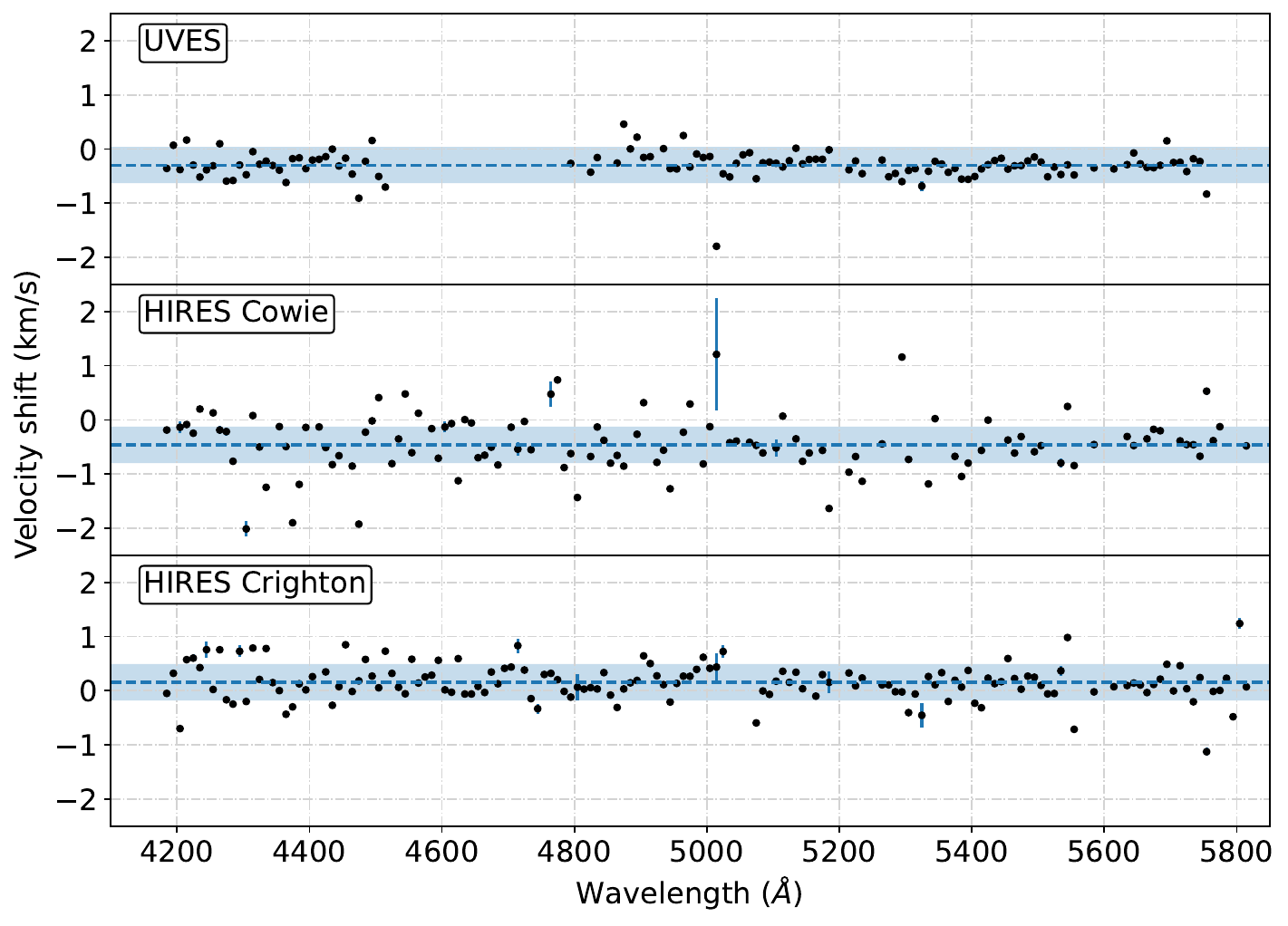}
    \caption{Velocity shift between UVES, the two HIRES and ESPRESSO spectra as a function of wavelength. 
    The error-bars on the y-axis represent the scatter in the determination of the velocity shift when computing the cross-correlation from 10 slightly different starting velocity values, equally spaced within a pixel. The adopted shifts, assumed to be constant, are $|v_\mathrm{shift}| = 0.295$ km s$^{-1}$, $0.463$ km s$^{-1}$ and $0.154$ km s$^{-1}$ respectively.}
    \label{fig:shift}
\end{figure}

The shifts between UVES and ESPRESSO depend weakly on the wavelength (Fig. \ref{fig:shift}). We therefore applied a single shift across the entire wavelength range compared, taken as the weighted average of the individual cross-correlation values we computed for the 10\AA{} window $|v_\mathrm{shift}| = 0.295$ km s$^{-1}$. This shift was applied to the UVES data as a fixed parameter in the VPFIT input file.

\subsection{Joint fit of UVES and ESPRESSO data}
Having determined the wavelength scale corrections for the UVES spectrum, we checked whether the previously developed model to describe ESPRESSO data only also describes the UVES spectrum well. We thus obtained $\log{\rm N_{H_I}} = \num{17.922\pm0.026}$ and $\log{\rm N_{D_I}} = \num{13.338\pm0.012}$, well in agreement with the result based on the ESPRESSO spectrum alone, although with a slightly lower hydrogen column density. The corresponding \ion{D}{I}/\ion{H}{I} is $\num{2.606\pm0.172e-5}$. 

We also performed a joint fit of the ESPRESSO and the corrected UVES data as VPFIT allows different data sets to be fit at the same time, using the same model, without having to resort to stacking or data manipulation techniques. The column densities derived from the joint fit are $\log{\rm N_{H_{I}}} = \num{17.922 \pm 0.012}$ and $\log{\rm N_{D_I}} = \num{13.338 \pm 0.012}$, leading to \ion{D}{I}/\ion{H}{I}  of $\num{2.608 \pm 0.102e-5}$.

Fixing the velocity shift makes it so that the uncertainty in the determination of the shift itself is not propagated to the \ion{D}{I}/\ion{H}{I} ratio. To estimate the uncertainty associated with this decision, we rerun the joint fit, but now fixing the shift to $-0.295 \pm 0.334$, that is to $0.039$ and $-0.6290$ \kms{} (the central value $\pm 1\sigma$). The resulting \ion{D}{I}/\ion{H}{I} associated with these shifts between UVES and ESPRESSO are, respectively, \num{2.612 \pm 0.102e-5} and \num{2.599 \pm 0.101e-5}. The statistical uncertainty dominates, and including the systematic uncertainty from the velocity shift increases the error on the final determination by about 1\%.

\subsection{Joint fit of ESPRESSO, UVES and HIRES data}
\label{sect:modelOnHIRES}
The HIRES data were corrected in an analogous way as the UVES data, after which we performed a joint fit of the ESPRESSO, UVES, and HIRES data. Interestingly, despite the increase in S/N achieved by including additional data ($\sim10$\% compared to the combination of ESPRESSO and UVES data), we find slightly larger errors in the $\log{\rm N_{H_{I}}}$ and $\log{\rm N_{D_{I}}}$ compared to the joint fit in ESPRESSO and UVES: $\log{\rm N_{H_I}} = \num{17.928 \pm 0.019}$, $\log{\rm N_{D_I}} = \num{13.343 \pm 0.017}$, \ion{D}{I}/\ion{H}{I} $ = \num{2.595\pm0.152e-5}$. The higher uncertainty can be attributed to several factors, including different normalisations for each spectra or region-dependent wavelength shifts (see Sect. \ref{sec:errorDiscussion}).
Because of this, we consider only the combined ESPRESSO + UVES data henceforth.

\section{Cloudy model of the system}
\label{sec:cloudyModel}
We use Cloudy \citep{cloudy_17} to estimate the metallicity of the system, the temperature of the absorber and the hydrogen density. Cloudy is a spectral synthesis code designed to simulate astrophysical environments and predict their spectra.
Given the presence of several transitions with different ionisation stages in the spectrum, we use the \texttt{optimize} function to estimate the quantities mentioned above.

We assume the gas to be in photoionisation equilibrium with the ultraviolet ionising background. We chose the HM12 option, i.e.\ an isotropic ionising background with the contribution of both quasars and galaxies, with a variable escape fraction $\langle f_{\rm esc}\rangle$ depending on the redshift of interest \citep{HM12}. The contribution of the CMB, although negligible, is also taken into account. We approximate the absorption systems as a plane parallel slab with hydrogen density $n_{\rm H}$ and metallicity $Z/Z_\odot$: these two quantities are free parameters, and the best-fit value was estimated by Cloudy during the optimisation process. We allowed the optimiser to explore a reasonably large range for both quantities: the allowed interval for the hydrogen density is $-5 < \log{(n_H)} < 2$ ${\rm cm^{-3}}$, for metallicity $-3 < \log{(Z/Z_\odot)} < 0$. We used the median point as initial guesses for each interval. Both initial guesses and the allowed ranges are consistent with values found in the literature.

Table \ref{tab:cloudyRes} reports the Cloudy output. The simulated column densities are generally consistent with those estimated by the fit. The most obvious exception to this is \ion{Fe}{III}, which appears to have significantly higher column density in our VPFIT model. This can be explained by the fact that the \ion{Fe}{III} lines lie in the Lyman forest and, as such, might be affected by Ly$\alpha$ interlopers or imperfect continuum placement.

\begin{table}
    \caption{Results of the optimisation process from Cloudy, and inferred column densities from the modelling. Metal species marked with * are formally upper limits, either because column densities were determined by force-fitting absorption lines at the position of the absorber (in the case of \ion{O}{I}), or the lines are in the Lyman forest, and as such a part of the absorption can be attributed to Ly$\alpha$ interlopers (everything else).}
    \centering
    \begin{tabular}{c|c|c}
        \toprule
        Metal species & Cloudy [cm$^{-2}$] & Fit [cm$^{-2}$] \\
        \midrule
        \ion{C}{II}   & 13.38  & \num{13.35 \pm 0.06}  \\
        \ion{C}{III}*  & 14.44 & <14.25  \\
        \ion{C}{IV}   & 12.88  & \num{13.01 \pm 0.14}  \\
        \ion{Si}{II}  & 12.45  & \num{12.42 \pm 0.07}  \\
        \ion{Si}{III}* & 13.53 & <13.90  \\
        \ion{Si}{IV}  & 12.79  & \num{13.10 \pm 0.10}  \\
        \ion{Fe}{III}* & 12.80 & <13.33  \\
        \ion{O}{I}* & 11.23    & <12.10  \\
        \bottomrule
    \end{tabular} \\

    \begin{tabular}{c|c}
        \toprule
        $n_H\ {\rm [cm^{-3}]}$ & -2.49  \\
        $Z/Z_\odot$ & -2.45  \\
        \bottomrule
    \end{tabular}

    \label{tab:cloudyRes}
\end{table}

The estimated hydrogen density for the cloud, $n_H$, is lower (by a factor of 3) than the minimum value determined by \citetalias{RS_17}, who compared the column density ratios observed with a grid of Cloudy models and estimated it to be $-2.11 < n_H < -1.72$ ${\rm cm^{-3}}$. The best estimate of metallicity is consistent with previous determinations ($-2.5 < Z/Z_\odot < -1.99$, \citetalias{RS_17}) and recent analysis of low-metallicity systems \citep{MAGG_23}. We note that our simulation uses an updated UV background with respect to \citetalias{RS_17}: re-running the model using the HM05 UV background, same as \citetalias{RS_17}, yields $n_H = -2.27 \ {\rm cm^{-3}}$, closer to the lower limit found by \citetalias{RS_17}. 
Finally, we obtain an average temperature for the system of $1.76\times 10^4 {\rm K}$, in good agreement with the temperature estimated by VPFIT.

\section{Discussion}
\label{sec:comparisonWithOtherModels}
Inferring the \ion{H}{I} column density is an inherently difficult task and various authors have reported different values for this absorption system, with inconsistent conclusions. \citet{tytler_96} found a value of $\log{\rm N_{H_I}} = 17.94 \pm 0.3 \pm 0.3$, where the first error is statistical and the second is systematic. The estimate was subsequently revised to $\log{\rm N_{H_I}} = 17.86 \pm 0.02$ cm$^{-2}$ based on the LRIS, HIRES and Kast data \citep{burles_tytler_hydrogen}. Improvement in the determination was achieved by developing a new method to estimate the \ion{H}{I} column density from the Lyman continuum optical depth. This result was however inconsistent with results from \citet[$\log{N_{\rm H_I}}<17.7$ cm$^{-2}$]{songaila_97} or \citet{wampler_96}, who again estimated a lower $\log{N_{\rm H_I}}$ column density.

In addition to measurements of the \ion{H}{I} column density, PKS1937-101 has three previous determinations of the D/H ratio \citep{tytler_96, burles_tytler_98, RS_17}. \citet{tytler_96} estimated a \ion{D}{}/\ion{H}{} ratio of $\num{2.3 \pm .3e-5}$, while improvements in the determination of \ion{H}{I} from \citet{burles_tytler_98} lead to an updated value of $\num{3.3 \pm .3e-5}$. However, in their analysis, metal components were not used to better constrain the velocity structure of the system. Instead, the position and the number of components for \ion{D}{I} and \ion{H}{I} are only constrained using Lyman lines. The discrepancy between their final determination and the value presented in this paper is due to the different \ion{H}{I} column density, lower in \citet{burles_tytler_98} than the present measurement (Tab. \ref{tab:DoH_Summary}).

More recently, the system has been revisited by \citetalias{RS_17}, who used data from Keck/LRIS, Keck/HIRES and VLT/UVES. They report a value of the primordial \ion{D}{I}/\ion{H}{I} of \num{2.62 \pm 0.051e-5}. Their determination of D/H is consistent and within $1\sigma$ of the one presented in this paper, albeit with smaller uncertainties. There are, however, some differences: our model does not include high-ionisation metal species, which might not faithfully trace the \ion{H}{I} velocity structure, leading to a three-component model (Sect. \ref{sect:modelFourComponents} shows how including a fourth component is only required to adequately fit the \ion{Si}{IV} doublet). In addition to this, in this work interlopers are modelled, when possible, as Ly$\beta$ or Ly$\gamma$ lines with an associated Ly$\alpha$ line at a longer wavelength, instead of simple Ly$\alpha$; this should make the model more robust, as interlopers are constrained more precisely.
We also note that the determination from the ESPRESSO data, even when combined with UVES, leads to a slightly larger uncertainty than \citetalias{RS_17}. We show that systematic uncertainties are significant (Sect. \ref{sect:model_unc}), and as such that the errors in previous analyses might be underestimated. 

\begin{table*}
    \caption{Summary of the \ion{D}{I}/\ion{H}{I} measurements for the system at $z=3.572$ in PKS 1937-101 quoted in the text.}
    \centering
    \begin{tabular}{lccccc}
    \toprule
        Publication                            & $\log{\rm N_{D_I}}$ & $\log{\rm N_{H_I}}$ & \ion{D}{I}/\ion{H}{I} (x 10$^{-5}$) & \# of components  & Res. power \\
    \midrule
        \citet{tytler_96}                      & \num{13.30    \pm 0.04}    & \num{17.94  \pm 0.05}   & \num{2.3   \pm 0.6}   & 2  & 37 000 \\
        \citet{burles_tytler_98}               & ///                        & \num{17.86  \pm 0.02}   & \num{3.24  \pm 0.30}  & 2  & 37 000 \\
        \citetalias{RS_17} (UVES)              & \num{13.332 \pm 0.0287}    & \num{17.925 \pm 0.0063} & \num{2.58  \pm 0.175} & 4  & 45 000 \\
        \citetalias{RS_17} (HIRES)             & \num{13.357 \pm 0.0244}    & \num{17.925 \pm 0.0066} & \num{2.70  \pm 0.157} & 4  & 37000 - 49000 \\
        \citetalias{RS_17} (Combined)          & \num{13.344 \pm 0.0056}    & \num{17.921 \pm 0.0068} & \num{2.62  \pm 0.051} & 4  & 37000 - 49000 \\
        This work (ESPRESSO)                   & \num{13.345 \pm 0.015}     & \num{17.923 \pm 0.015}  & \num{2.638 \pm 0.128} & 3  & 70 000 \\
        This work (UVES)                       & \num{13.338 \pm 0.012}     & \num{17.923 \pm 0.026}  & \num{2.60  \pm 0.17}  & 3  & 45 000 \\
        This work (HIRES)                      & \num{13.357 \pm 0.029}     & \num{17.895 \pm 0.019}  & \num{2.89  \pm 0.23}  & 3  & 37000 - 49000 \\
        This work (ESPRESSO + UVES)            & \num{13.338 \pm 0.012}     & \num{17.922 \pm 0.012}  & \num{2.608 \pm 0.102} & 3  & --- \\
        This work (ESPRESSO + HIRES)           & \num{13.353 \pm 0.0079}    & \num{17.910 \pm 0.013}  & \num{2.770 \pm 0.094}    & 3  & --- \\
        This work (ESPRESSO + UVES + HIRES)    & \num{13.343 \pm 0.017}     & \num{17.928 \pm 0.019}  & \num{2.595 \pm 0.152} & 3  & --- \\
    \bottomrule
    \end{tabular}

    \label{tab:DoH_Summary}
\end{table*}

\subsection{Determinations of \texorpdfstring{\ion{D}{}/\ion{H}{}}{} from other systems in the literature.}
\begin{figure}
    \centering
    \includegraphics[width=\columnwidth]{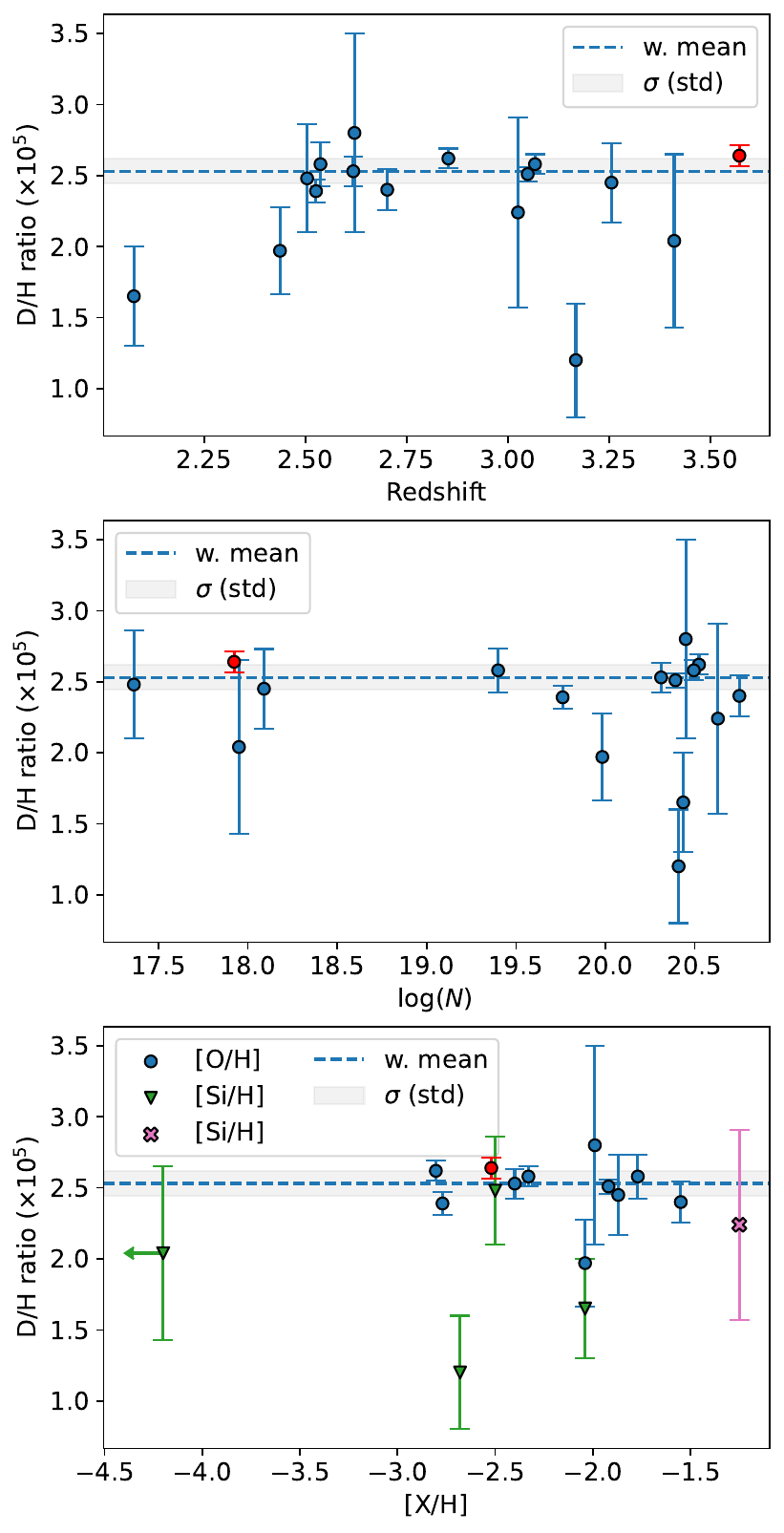}
    \caption{Data from Tab. \ref{tab:otherDeuterium}, plotted from top to bottom against redshift, \ion{H}{I} column density and metallicity of the absorber. The red point is the new measurement reported in this work, the grey, dashed line the weighted mean of measurements in Tab. \ref{tab:otherDeuterium}, while the grey shaded regions represent the $1\sigma$ standard deviation. In the bottom panel, metallicity is estimated through different metal ratios, colour coded according to the legend on the top left. No clear correlations are evident.}
    \label{fig:comparison_other_systems}
\end{figure}

We compare the determination presented in this work against the determinations from other systems, including the Precision Sample defined by \citet{cooke_18}. 
In Table \ref{tab:otherDeuterium}, we report the "robust" \ion{D}{}/\ion{H}{} measurements initially selected by \citet{pettini_2008} and updated in subsequent articles (RS15, \citetalias{RS_17}, \citet{zavarygin_18}) with the new determination derived in this paper for the system at z=3.572 in PKS1937-101. Values from the same table are plotted in Fig. \ref{fig:comparison_other_systems}.

\begin{table*}
    \centering
    \begin{tabular}{l|l|c|c|c|c}
        \toprule
         Reference & Quasar & redshift of the system & $\log{\rm N_{H_I}}$ & [X/H] & \ion{D}{}/\ion{H}{} (x 10$^5$) \\
         \midrule
         \citet{pettini_2001}           & Q2206-199       & 2.076  & \num{20.436 \pm 0.008}   & -2.04 [Si/H]   & \num{1.65 \pm 0.35} \\
         \citet{balashev_2016}          & J1444+2919      & 2.437  & \num{19.983 \pm 0.010}   & -2.04 [O/H]    & \num{1.97 \pm 0.33} \\
         \citet{zavarygin_18}           & Q1009+2956      & 2.504  & \num{17.362 \pm 0.005}   & -2.50 [Si/H]   & \num{2.48 \pm 0.41} \\
         \citet{cooke_18}*              & Q1243+3047      & 2.525  & \num{19.761 \pm 0.026}   & -2.77 [O/H]    & \num{2.39 \pm 0.10} \\
         \citet{omeara_2001}*           & HS 0105+1619    & 2.536  & \num{19.40  \pm 0.01}    & -1.77 [O/H]    & \num{2.58 \pm 0.15} \\
         \citet{pettini_2008}*          & Q0913+072       & 2.618  & \num{20.312 \pm 0.008}   & -2.40 [O/H]    & \num{2.53 \pm 0.10} \\
         \citet{noterdaeme_2012}        & J0407-4410      & 2.621  & \num{20.45  \pm 0.10}    & -1.99 [O/H]    & \num{2.80 \pm 0.8} \\
         \citet{omeara_2006}*           & J1558-0031      & 2.702  & \num{20.75  \pm 0.03}    & -1.55 [O/H]    & \num{2.40 \pm 0.14} \\
         \citet{cooke_16}*              & SDSS J1358+0349 & 2.853  & \num{20.524 \pm 0.006}   & -2.804 [O/H]   & \num{2.62 \pm 0.07} \\
         \citet{dodorico_01}            & QSO 0347-3819   & 3.025  & \num{20.63  \pm 0.09}    & -1.25 [Zn/H]   & \num{2.24 \pm 0.67} \\
         \citet{pettini_2012}*          & J1419+0829      & 3.049  & \num{20.392 \pm 0.003}   & -1.92 [O/H]    & \num{2.51 \pm 0.05} \\
         \citet{cooke_14}*              & J1358+6522      & 3.067  & \num{20.50  \pm 0.01}    & -2.33 [O/H]    & \num{2.58 \pm 0.10} \\
         \citet{Srianand_10}            & J1337+3152      & 3.168  & \num{20.41  \pm 0.15}    & -2.68 [Si/H]   & \num{1.2  \pm 0.5} \\
         \citet{RS_15}                  & Q1937-101       & 3.256  & \num{18.09  \pm 0.03}    & -1.87 [O/H]    & \num{2.45 \pm 0.28} \\
         \citet{fumagalli_2011}         & J1134+5742      & 3.411  & \num{17.95  \pm 0.05}    & < -4.20 [Si/H] & \num{2.04 \pm 0.61} \\
         This work                      & PKS1937-101       & 3.572  & \num{17.922 \pm 0.023}   & -2.52 [O/H]  & \num{2.608 \pm 0.102} \\
         \bottomrule
    \end{tabular}
    \caption{Collection of measurements of the primordial deuterium abundance from the literature. Starred QSOs are part of the Precision Sample presented in \citet{cooke_18}.}
    \label{tab:otherDeuterium}
\end{table*}

Qualitatively, there is no apparent correlation between redshift, metallicity, column density of the absorber and the \ion{D}{}/\ion{H}{} ratio. In particular, the absence of correlation with redshift strengthens the result by \citet{cooke_18} since the updated sample spans a wider redshift range. 

The collection presented in this section is inhomogeneous and includes results obtained with different hypotheses (for example, \citet{noterdaeme_2012, balashev_2016} assume a constant ratio of \ion{O}{I}/\ion{H}{I} across all components of the system, \citet{cooke_18} allow the temperature of hydrogen and deuterium to vary independently). Averaging all the measurements reported in Tab. \ref{tab:otherDeuterium}, we find simple and weighted averages of \num{2.32 \pm 0.4e-5} and \num{2.512\pm 0.028e-5}, respectively. As expected, the weighted average is consistent with the results of \citet{cooke_18}, as 7 out of 16 measurements in Tab. \ref{tab:otherDeuterium} are also in the Precision Sample. These also have the smallest formal errors, so it is expected that the weighted average of all 16 measurements in Tab. \ref{tab:otherDeuterium} is close to the Precision Sample average. The simple average, on the other hand, is lower and inconsistent with the value provided by the Precision Sample at the level of $\sim2 \sigma$. This is due to the measurements with very discrepant lower values, which, however, also have larger error bars, and in general to the fact that the sample has not been analysed homogeneously. Taking into account only measurements with an error estimate lower than 10\% the simple mean results in \num{2.53 \pm 0.09e-5}, while the weighted average is almost unchanged \num{2.531 \pm 0.029e-5}. On the other hand, comparing against the Precision Sample, for which \citet{cooke_18} report a determination of the primordial deuterium abundance of \num{2.527\pm0.030e-5}, we find that the result from ESPRESSO data alone is in agreement as is the one obtained with the combination of ESPRESSO and UVES data, at the $\sim0.8\sigma$ level.  

\begin{figure}
    \centering
    \includegraphics[width=\columnwidth]{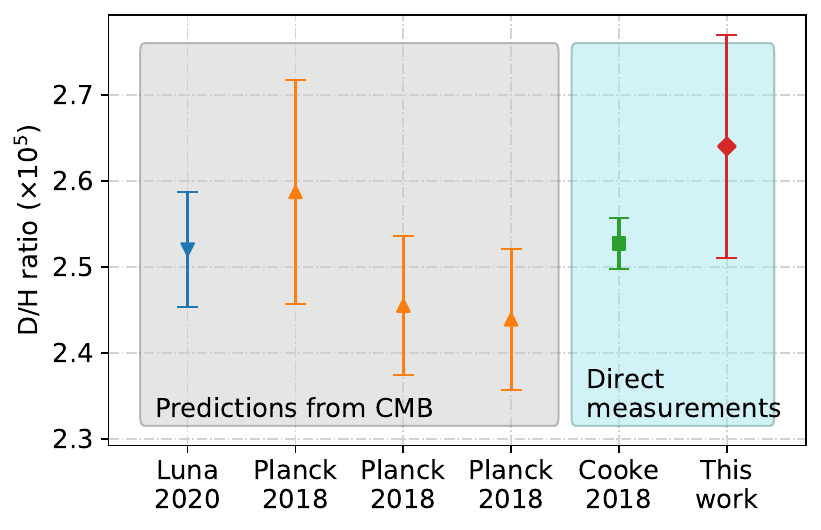}
    \caption{Primordial deuterium abundance from the LUNA experiment \citep{luna_deuterium}, Planck data and the Precision Sample presented in \citet{cooke_18}, compared to the results of the present work. Planck provides three different measurements, depending on the assumed nuclear rates \citep{Planck_2018_cosmo_pars}.}
    \label{fig:Planck_Luca_Cooke}
\end{figure}

Finally, we compare the new determination with the most recent results from the Planck Collaboration \citep{Planck_2018_cosmo_pars}. Converting CMB data to deuterium abundance requires prior knowledge of nuclear reaction rates, and the Planck Collaboration presents three possibilities (\ion{D}{}/\ion{H}{} = \num{2.587 \pm 0.13}, \num{2.455 \pm 0.081}, \num{2.439 \pm 0.082}), depending on the nuclear reaction rates considered. The combined ESPRESSO and UVES measurement agrees with these results at $0.2\sigma,\ 1.2\sigma$ and $1.3\sigma$, respectively (Fig. \ref{fig:comparison_other_systems}). More recently, the LUNA experiment updated the cross-section of the deuterium burning D($p,\gamma)^3$He reaction \citep{luna_deuterium} and provided a new determination of \ion{D}{I}/\ion{H}{I} of $2.52\ \pm 0.03\ \pm\ 0.06 \times 10^{-5}$, in agreement with the determination of this work at the $0.7\sigma$ level (Fig. \ref{fig:Planck_Luca_Cooke}). Thus, no significant tension is found, even if the offset in the central value suggests that systematic uncertainties for this system might be underestimated.

\subsection{About the uncertainty}
\label{sec:errorDiscussion}
The results obtained in this analysis have a higher uncertainty than \citetalias{RS_17}. This is somewhat unexpected, given the higher resolution, S/N, and significantly better wavelength calibration of ESPRESSO. We attribute this to two aspects: on the one hand, we attempt to fit together spectra from 3 different instruments. On the other hand, comparing the joint UVES and HIRES results with those obtained on single spectra, we recognise an unusual scaling in the final uncertainty reported by \citetalias{RS_17}.

With respect to the first point, UVES and HIRES have been shown in the past to be affected by systematics in the wavelength calibration. To account for this, we precompute the velocity shift of each fitting region and use the average value as a fixed parameter during the optimisation procedure. This is somewhat unique to this work: in \citetalias{RS_17}, for instance, a velocity shift between regions is allowed as a free parameter. In addition to this, each spectrum is reduced following slightly different prescriptions, and the continuum is determined with different algorithms. Systematic effects are thus expected and are hard to control. We argue that the increase in the uncertainty in the final measurement simply reflects the increased incompatibility of the input spectra. In support of this, fitting the model on HIRES data alone yields higher (albeit still consistent) \ion{D}{I}/\ion{H}{I} than ESPRESSO, UVES or the combination of the two (\ion{D}{I}/\ion{H}{I}$_{\rm HIRES}$ = \num{2.89 \pm 0.23e-5}).

In addition to this, we also note an atypical trend in the scaling of the final uncertainties reported by \citetalias{RS_17}: the uncertainty when fitting the model on single spectra (either UVES or HIRES) is a factor of three larger than the corresponding value found on the combined fit. Instead, we would expect the error to scale with the square root of S/N. This would lead to an uncertainty on the final determination by \citetalias{RS_17} of $\sim0.11$: systematics would further increase this number. This is quantified in Sect. \ref{sec:ContinuumPlacement} and Sect. \ref{sec:fittingRegion}. We find that, for example, the choice of a narrower fitting region combined with continuum adjustments shifts the central value of the \ion{D}{I}/\ion{H}{I} by nearly one $\sigma$. Based on these arguments, we show that the uncertainty of the determination is higher than previously reported and is likely dominated by systematics rather than statistical uncertainty.

\section{Summary and conclusions}
We have presented a new determination of the abundance of primordial deuterium in an absorption system at redshift $z=3.572$ seen towards the quasar PKS1937-101. The system was already studied in the past with different instruments \citep{tytler_96, burles_tytler_98, RS_17}. It is a Lyman limit system ($\log{\rm N_{H_I} \sim 17.9}$) showing a clear absorption due to \ion{D}{I} at the redshift of the \ion{H}{I} Ly$\alpha$ and Lyman $\beta$ lines. The analysis was carried out using new ESPRESSO data, of higher quality compared to previous measurements based on UVES and HIRES data alone. The new ESPRESSO data have 50\% higher S/N per \kms{} at almost double spectral resolution and significantly better wavelength calibration. In addition to the analysis of the ESPRESSO data alone, archival data have been included to further reduce the final measurement uncertainties.

Based on ESPRESSO data alone, we found a \ion{D}{I}/\ion{H}{I} of \num{2.638 \pm 0.128e-5} (Sect. \ref{sect:DoH_ESPRESSO}). This is slightly higher than the most recent measurement by \citetalias{RS_17}, although consistent within the respective error bars.
Including archival UVES data, we found \ion{D}{}/\ion{H}{} = \num{2.608 \pm 0.102e-5} (Sect. \ref{sect:modelOnUVES}), while the inclusion of HIRES data did not decrease the uncertainty, possibly indicating that we have reached the limit of statistical errors (Sect. \ref{sect:modelOnHIRES} and Sect. \ref{sec:errorDiscussion}). The systematic effects that may explain this are difficult to model.
The determination of the joint ESPRESSO+UVES fit is consistent with the weighted mean presented by \citet{cooke_18} based on the Precision Sample ($\sim 0.8 \sigma$ level), Planck data \citep[$0.2, 1.2, 1.3 \sigma$ level,][]{Planck_2018_cosmo_pars}, and recent results from the LUNA experiment \citep[$\sim 0.7 \sigma$ level,][]{luna_deuterium}.

Using ESPRESSO data, we discussed uncertainties in the modelling. We quantified the impact of continuum estimation uncertainties on the final measurement and investigated whether a fourth component is needed, finding that it is only required to adequately fit higher ionisation lines, but it is removed when considering low-ionisation metals, \ion{D}{I} and \ion{H}{I}. Furthermore, we found that choosing a smaller fitting region for the Ly$\alpha$ region leads to a higher determination of \ion{D}{}/\ion{H}{} of \num{2.76\pm0.16e-5}: this is due to VPFIT adjusting the continuum to be lower, reducing the \ion{H}{I} column density. Our model also fits well the higher-order Lyman series lines.

Surprisingly, the final determination from the ESPRESSO+UVES data carries higher uncertainties than those reported by \citetalias{RS_17}.
Column densities of \ion{H}{I} lines in the Lyman limit range ($17.3 \le \log{\rm N_{H_I}} \le 19$), as the one analysed here, are hard to constrain precisely because the line is saturated but it is not yet showing the Damped wings of the Lorentzian profile. This is evidenced by the disagreement among independent determinations. Moreover, systematic errors from the continuum placement or the choice of the fitting region are significant.

Finally, we combine the results of this work with a collection of results from the literature. A weighted average of the sample leads to \ion{D}{}/\ion{H}{}=\num{2.512\pm 0.028e-5}. This changes to \ion{D}{}/\ion{H}{}=\num{2.531\pm 0.029e-5} if we exclude measurements with uncertainties greater than 10\%. Considering the whole sample, we do not find any correlation between the \ion{D}{}/\ion{H}{} ratio and the redshift, metallicity or the \ion{H}{I} column density, confirming the findings of \citet{cooke_18}. The results of the combined sample are also consistent with both Planck \citep{Planck_2018_cosmo_pars} and LUNA \citep{luna_deuterium} experiments: no significant tension is evident.

\section*{Acknowledgements}
The authors acknowledge the ESPRESSO project team for its effort and dedication in building the ESPRESSO instrument.
The INAF authors acknowledge the financial support of the Italian Ministry of Education, University, and Research with PRIN 201278X4FL and the "Progetti Premiali" funding scheme. 
SC and DM are partly supported by the INFN PD51 INDARK grant.
JIGH, ASM, CAP and RR acknowledge financial support from the Spanish Ministry of Science and Innovation (MICINN) project PID2020-117493GB-I00.
This work was supported by Portuguese funds through FCT - Funda\c c\~ao para a Ci\^encia e a Tecnologia through national funds and in the framework of the project 2022.04048.PTDC (Phi in the Sky, DOI 10.54499/2022.04048.PTDC), 
and by FEDER through COMPETE2020 - Programa Operacional Competitividade e Internacionalização by these grants: UIDB/04434/2020; UIDP/04434/2020.
CJM also acknowledges FCT and POCH/FSE (EC) support through Investigador FCT Contract 2021.01214.CEECIND/CP1658/CT0001. 
S.G.S acknowledges the support from FCT through Investigador FCT contract nr. CEECIND/00826/2018 and  POPH/FSE (EC).
FPE would like to acknowledge the Swiss National Science Foundation (SNSF) for supporting research with ESPRESSO through the SNSF grants nr. 140649, 152721, 166227 and 184618. The ESPRESSO Instrument Project was partially funded through SNSF’s FLARE Programme for large infrastructures. 
TMS acknowledgment the support from the SNF synergia grant CRSII5-193689 (BLUVES)
MTM acknowledges the support of the Australian Research Council through Future Fellowship grant FT180100194.

We acknowledge financial support from the Agencia Estatal de Investigaci\'on of the Ministerio de Ciencia e Innovaci\'on MCIN/AEI/10.13039/501100011033 and the ERDF “A way of making Europe” through project PID2021-125627OB-C32, and from the Centre of Excellence “Severo Ochoa” award to the Instituto de Astrofisica de Canarias.

\section*{Data Availability}
The data underlying this article will be shared on reasonable request to the corresponding author.

\bibliographystyle{mnras}
\bibliography{bib} %

\appendix
\onecolumn

\input{Model}
\input{TransitionUnrelated}

\bsp	%
\label{lastpage}
\end{document}

%% file: Model.tex
\section{VPFIT model}

\begin{longtable}{ccccccccccc}
    \caption{We report for each transition included in the model the corresponding column density, redshift and total $b$ parameter. For the transitions where we fit for both $b_{\rm turb}$ and T we report the error separately for both quantities, and not on the b$_{\rm tot}$. For transitions marked with *, the reported column density is the sum of the column densities across the three components, and only report the error on the sum.} \\
    \label{tab:modelTransition} \\
   \toprule
   Transition & $\log({\rm N})$ & $\Delta \log({\rm N})$ & $z$ & $\Delta z$ &  $b_{\rm tot}$ & $\Delta b_{\rm tot}$ & $b_{\rm turb}$ & $\Delta b_{\rm turb}$ & $T$ & $\Delta T$ \\
    & & & & & [\kms{}] & [\kms{}] & [\kms{}] & [\kms{}] & [K] & [K] \\ 
   \midrule
        \ion{C}{II}    &  12.27719  &  0.09592  &  3.5721324  &  0.0000112  &  5.10710  &   ////   &  2.01  &  2.15  &  1590  &  459  \\
        \ion{C}{II}    &  12.80011  &  0.05103  &  3.5722725  &  0.0000050  &  6.72580  &   ////   &  4.76  &  0.87  &  1630  &  5190 \\
        \ion{C}{II}    &  13.14995  &  0.01827  &  3.5724591  &  0.0000036  &  7.63190  &   ////   &  5.72  &  0.32  &  1840  &  1110 \\
        \ion{Si}{II}   &  11.07122  &  0.16452  &  3.5721324  &  0.0000000  &  3.67030  &   ////   &  2.01  &  2.15  &  1590  &  459  \\
        \ion{Si}{II}   &  11.92929  &  0.04499  &  3.5722725  &  0.0000000  &  5.68360  &   ////   &  4.76  &  0.87  &  1630  &  5190 \\
        \ion{Si}{II}   &  12.22167  &  0.01869  &  3.5724591  &  0.0000000  &  6.60520  &   ////   &  5.72  &  0.32  &  1840  &  1110 \\
        \ion{Fe}{III}  &  12.68527  &  0.08734  &  3.5722725  &  0.0000000  &  5.22720  &   ////   &  ////  &  ////  &  ////  &  //// \\
        \ion{Fe}{III}  &  13.20000  &  0.02378  &  3.5724591  &  0.0000000  &  6.16460  &   ////   &  ////  &  ////  &  ////  &  //// \\
        \ion{H}{I}     &  13.49750  &  0.01004  &  3.2179784  &  0.0000096  &  45.4376  &  0.9483  &  ////  &  ////  &  ////  &  //// \\
        \ion{H}{I}     &  13.17590  &  0.31649  &  3.2192094  &  0.0002257  &  33.3361  &  8.9871  &  ////  &  ////  &  ////  &  //// \\
        \ion{H}{I}     &  13.75953  &  0.07837  &  3.2195409  &  0.0000044  &  20.8896  &  1.0158  &  ////  &  ////  &  ////  &  //// \\
        \ion{H}{I}     &  14.66222  &  0.08622  &  3.2209637  &  0.0000079  &  28.3678  &  0.7715  &  ////  &  ////  &  ////  &  //// \\
        \ion{H}{I}     &  14.92678  &  0.05668  &  3.2210411  &  0.0000097  &  19.5230  &  1.0635  &  ////  &  ////  &  ////  &  //// \\
        \ion{H}{I}     &  14.37814  &  0.01210  &  3.2238886  &  0.0000066  &  25.1154  &  0.3731  &  ////  &  ////  &  ////  &  //// \\
        \ion{H}{I}     &  13.51177  &  0.00527  &  3.2256770  &  0.0000037  &  26.5957  &  0.3201  &  ////  &  ////  &  ////  &  //// \\
        \ion{H}{I}     &  14.13789  &  0.01365  &  3.2246176  &  0.0000099  &  27.8596  &  0.5640  &  ////  &  ////  &  ////  &  //// \\
        \ion{H}{I}     &  11.90997  &  0.09105  &  3.2228270  &  0.0000323  &  16.4572  &  3.6905  &  ////  &  ////  &  ////  &  //// \\
        \ion{H}{I}     &  12.75313  &  0.12461  &  3.4818366  &  0.0000234  &  17.9989  &  2.0488  &  ////  &  ////  &  ////  &  //// \\
        \ion{H}{I}     &  13.55302  &  0.11481  &  3.4826829  &  0.0000083  &  24.7134  &  2.0620  &  ////  &  ////  &  ////  &  //// \\
        \ion{H}{I}     &  13.75280  &  0.19026  &  3.4833953  &  0.0000726  &  55.8278  &  6.6979  &  ////  &  ////  &  ////  &  //// \\
        \ion{H}{I}     &  12.55664  &  0.08984  &  3.4836069  &  0.0000130  &  13.1128  &  1.5776  &  ////  &  ////  &  ////  &  //// \\
        \ion{H}{I}     &  13.22525  &  0.40686  &  3.4847351  &  0.0004795  &  59.1222  &  9.0805  &  ////  &  ////  &  ////  &  //// \\
        \ion{H}{I}     &  14.46741  &  0.00494  &  3.4860559  &  0.0000021  &  27.4336  &  0.1766  &  ////  &  ////  &  ////  &  //// \\
        \ion{H}{I}     &  12.77936  &  0.13431  &  3.4871438  &  0.0000120  &  23.9338  &  2.9068  &  ////  &  ////  &  ////  &  //// \\
        \ion{H}{I}     &  12.86475  &  0.15360  &  3.4878010  &  0.0002088  &  56.8514  &  3.9931  &  ////  &  ////  &  ////  &  //// \\
        \ion{H}{I}     &  12.14067  &  0.09033  &  3.4893585  &  0.0000291  &  20.4586  &  3.2330  &  ////  &  ////  &  ////  &  //// \\
        \ion{H}{I}     &  12.32798  &  0.08783  &  3.4903180  &  0.0000230  &  26.0675  &  3.4951  &  ////  &  ////  &  ////  &  //// \\
        \ion{H}{I}     &  17.92361  &  0.01547  &  3.5721324  &  0.0000000  &  16.3286  &  0.0000  &  ////  &  ////  &  ////  &  //// \\
        \ion{H}{I}     &  17.38427  &  0.11517  &  3.5722725  &  0.0000000  &  17.0856  &  0.0000  &  ////  &  ////  &  ////  &  //// \\
        \ion{H}{I}     &  17.46396  &  0.05434  &  3.5724591  &  0.0000000  &  18.3583  &  0.0000  &  ////  &  ////  &  ////  &  //// \\
        \ion{D}{I}     &  13.34481  &  0.01514  &  3.5721324  &  0.0000000  &  11.6781  &  0.0000  &  ////  &  ////  &  ////  &  //// \\
        \ion{D}{I}     &  12.80547  &  0.00000  &  3.5722725  &  0.0000000  &  12.5836  &  0.0000  &  ////  &  ////  &  ////  &  //// \\
        \ion{D}{I}     &  12.88516  &  0.00000  &  3.5724591  &  0.0000000  &  13.6409  &  0.0000  &  ////  &  ////  &  ////  &  //// \\
        \ion{H}{I}     &  13.42287  &  0.00717  &  3.5632977  &  0.0000053  &  32.3439  &  0.4975  &  ////  &  ////  &  ////  &  //// \\
        \ion{H}{I}     &  13.23902  &  2.29951  &  3.5643157  &  0.0005578  &  22.4179  &  9.8904  &  ////  &  ////  &  ////  &  //// \\
        \ion{H}{I}     &  13.56131  &  0.90470  &  3.5647841  &  0.0000404  &  14.5604  &  8.3881  &  ////  &  ////  &  ////  &  //// \\
        \ion{H}{I}     &  14.20576  &  0.48146  &  3.5649199  &  0.0003255  &  34.5775  &  7.9958  &  ////  &  ////  &  ////  &  //// \\
        \ion{H}{I}     &  13.16800  &  0.38599  &  3.5652351  &  0.0000398  &  11.3401  &  2.7810  &  ////  &  ////  &  ////  &  //// \\
        \ion{H}{I}     &  14.03812  &  0.01663  &  3.5663878  &  0.0000076  &  27.5262  &  0.5996  &  ////  &  ////  &  ////  &  //// \\
        \ion{H}{I}     &  13.55277  &  0.04721  &  3.5670652  &  0.0000420  &  35.2826  &  1.8478  &  ////  &  ////  &  ////  &  //// \\
        \ion{H}{I}     &  12.14010  &  0.06582  &  3.5683673  &  0.0000328  &  21.5655  &  3.4345  &  ////  &  ////  &  ////  &  //// \\
        \ion{H}{I}     &  13.04629  &  0.01142  &  3.5695084  &  0.0000068  &  31.4880  &  0.7736  &  ////  &  ////  &  ////  &  //// \\
        \ion{H}{I}     &  15.76831  &  0.70475  &  3.5728219  &  0.0004595  &  29.0785  &  6.4334  &  ////  &  ////  &  ////  &  //// \\
        \ion{H}{I}     &  13.72076  &  1.87345  &  3.5735586  &  0.0004771  &  21.5114  &  9.2989  &  ////  &  ////  &  ////  &  //// \\
        \ion{H}{I}     &  13.10764  &  0.04477  &  3.5750312  &  0.0000521  &  57.6792  &  5.5815  &  ////  &  ////  &  ////  &  //// \\
        \ion{H}{I}     &  12.23112  &  0.18153  &  3.5757240  &  0.0000160  &  17.2274  &  3.4219  &  ////  &  ////  &  ////  &  //// \\
        \ion{H}{I}     &  13.64665  &  0.00514  &  3.5766944  &  0.0000023  &  20.5472  &  0.2636  &  ////  &  ////  &  ////  &  //// \\
        \ion{H}{I}     &  12.57496  &  0.08553  &  3.5772601  &  0.0000156  &  12.8801  &  1.5762  &  ////  &  ////  &  ////  &  //// \\
        \ion{H}{I}     &  13.54550  &  0.01491  &  3.5777517  &  0.0000049  &  20.1200  &  0.7205  &  ////  &  ////  &  ////  &  //// \\
        \ion{H}{I}     &  13.60914  &  0.14873  &  3.5784697  &  0.0000411  &  19.7491  &  1.6802  &  ////  &  ////  &  ////  &  //// \\
        \ion{H}{I}     &  12.55723  &  1.63382  &  3.5788174  &  0.0007554  &  22.3322  &  3.6585  &  ////  &  ////  &  ////  &  //// \\
        \ion{H}{I}     &  12.23596  &  0.10544  &  3.5796144  &  0.0000286  &  16.4360  &  3.4158  &  ////  &  ////  &  ////  &  //// \\
        \ion{H}{I}     &  12.59238  &  0.03925  &  3.5801276  &  0.0000178  &  18.7243  &  1.4348  &  ////  &  ////  &  ////  &  //// \\
        \ion{H}{I}     &  13.22402  &  0.01686  &  3.5817311  &  0.0000062  &  29.2061  &  0.9144  &  ////  &  ////  &  ////  &  //// \\
        \ion{H}{I}     &  12.28541  &  0.16346  &  3.5817898  &  0.0000163  &  12.4031  &  2.2539  &  ////  &  ////  &  ////  &  //// \\
        \ion{H}{I}     &  12.61916  &  0.13766  &  2.6569922  &  0.0000857  &  36.5521  &  9.7131  &  ////  &  ////  &  ////  &  //// \\
        \ion{H}{I}     &  12.63054  &  0.17887  &  2.6606367  &  0.0000191  &  13.5772  &  4.1641  &  ////  &  ////  &  ////  &  //// \\
        \ion{H}{I}     &  13.33306  &  0.50016  &  2.6592406  &  0.0001951  &  21.1861  &  8.3306  &  ////  &  ////  &  ////  &  //// \\
        \ion{H}{I}     &  13.96424  &  0.43684  &  2.6596522  &  0.0000395  &  16.7285  &  5.9622  &  ////  &  ////  &  ////  &  //// \\
        \ion{H}{I}     &  14.12626  &  0.32432  &  2.6598944  &  0.0001896  &  30.2413  &  8.3448  &  ////  &  ////  &  ////  &  //// \\
        \ion{??}{}     &  12.92620  &  0.43025  &  2.6592894  &  0.0000184  &   7.8699  &  2.7932  &  ////  &  ////  &  ////  &  //// \\
        \ion{H}{I}     &  12.97916  &  0.03992  &  2.5698604  &  0.0000192  &  23.3382  &  1.9266  &  ////  &  ////  &  ////  &  //// \\
        \ion{H}{I}     &  12.80392  &  0.11076  &  2.5711063  &  0.0000337  &  14.6142  &  2.4086  &  ////  &  ////  &  ////  &  //// \\
        \ion{H}{I}     &  14.31257  &  0.07072  &  2.5738291  &  0.0000231  &  33.6546  &  4.5900  &  ////  &  ////  &  ////  &  //// \\
        \ion{H}{I}     &  13.57488  &  0.07094  &  2.5745386  &  0.0000196  &  19.3668  &  1.0655  &  ////  &  ////  &  ////  &  //// \\
        \ion{H}{I}     &  12.76783  &  0.20193  &  2.5729993  &  0.0000419  &  16.8660  &  6.2359  &  ////  &  ////  &  ////  &  //// \\
        \ion{H}{I}     &  13.03850  &  0.05278  &  2.5754521  &  0.0000227  &  25.1392  &  2.0029  &  ////  &  ////  &  ////  &  //// \\
        \ion{H}{I}     &  12.32696  &  0.42836  &  2.5257436  &  0.0000211  &   6.0693  &  2.0157  &  ////  &  ////  &  ////  &  //// \\
        \ion{H}{I}     &  12.39755  &  0.38378  &  2.5258665  &  0.0000804  &  10.0218  &  5.6534  &  ////  &  ////  &  ////  &  //// \\
        \ion{H}{I}     &  13.38328  &  0.08656  &  2.5257474  &  0.0000346  &  80.1206  &  1.0527  &  ////  &  ////  &  ////  &  //// \\
        \ion{H}{I}     &  13.33289  &  0.23233  &  2.5277668  &  0.0001658  &  36.8602  &  9.1620  &  ////  &  ////  &  ////  &  //// \\
        \ion{H}{I}     &  13.56498  &  0.04497  &  2.5296692  &  0.0000249  &  64.0627  &  4.3216  &  ////  &  ////  &  ////  &  //// \\
        \ion{H}{I}     &  13.57421  &  0.20761  &  2.4990371  &  0.0005172  &  73.5544  &  4.3406  &  ////  &  ////  &  ////  &  //// \\
        \ion{H}{I}     &  13.44619  &  0.09070  &  2.4994164  &  0.0000068  &  21.1994  &  1.8436  &  ////  &  ////  &  ////  &  //// \\
        \ion{??}{}     &  11.95068  &  0.12458  &  2.4829808  &  0.0000142  &   5.9523  &  2.3065  &  ////  &  ////  &  ////  &  //// \\
        \ion{H}{I}     &  12.62854  &  0.07857  &  2.4822349  &  0.0000206  &  27.8076  &  3.8507  &  ////  &  ////  &  ////  &  //// \\
        \ion{??}{}     &  11.98511  &  0.17310  &  2.4843733  &  0.0000110  &   2.3946  &  2.2016  &  ////  &  ////  &  ////  &  //// \\
        \ion{H}{I}     &  12.90598  &  0.22810  &  2.4868100  &  0.0000597  &  22.4348  &  8.4080  &  ////  &  ////  &  ////  &  //// \\
        \ion{H}{I}     &  13.41392  &  0.06346  &  2.4863433  &  0.0000332  &  26.4205  &  2.3093  &  ////  &  ////  &  ////  &  //// \\
        \ion{H}{I}     &  13.64106  &  0.14584  &  2.4848864  &  0.0000729  &  24.8313  &  4.3022  &  ////  &  ////  &  ////  &  //// \\
        \ion{H}{I}     &  13.48436  &  0.10874  &  2.4712119  &  0.0001014  &  47.0441  &  6.4405  &  ////  &  ////  &  ////  &  //// \\
        \ion{H}{I}     &  13.45318  &  0.15149  &  2.4715520  &  0.0000375  &  18.1789  &  3.3932  &  ////  &  ////  &  ////  &  //// \\
        \ion{H}{I}     &  13.61457  &  0.05130  &  2.4753132  &  0.0000179  &  21.8047  &  1.0605  &  ////  &  ////  &  ////  &  //// \\
        \ion{H}{I}     &  12.07801  &  0.13165  &  2.4734717  &  0.0000226  &  11.2243  &  3.6924  &  ////  &  ////  &  ////  &  //// \\
        \ion{H}{I}     &  13.16757  &  0.02739  &  2.4729291  &  0.0000122  &  26.8911  &  1.9385  &  ////  &  ////  &  ////  &  //// \\
        \ion{H}{I}     &  13.70755  &  0.04148  &  2.4749054  &  0.0000225  &  25.7499  &  1.3743  &  ////  &  ////  &  ////  &  //// \\
        \ion{H}{I}     &  12.41494  &  0.08154  &  2.4739942  &  0.0000204  &  19.0407  &  3.2312  &  ////  &  ////  &  ////  &  //// \\
        \ion{H}{I}     &  14.18948  &  0.03554  &  2.4632222  &  0.0000269  &  36.8891  &  1.4545  &  ////  &  ////  &  ////  &  //// \\
    \bottomrule
\end{longtable}

%% file: TransitionUnrelated.tex
\begin{longtable}{cccccccc}
    \caption{Interlopers that contribute as Lyman $\beta$ or Lyman $\gamma$ to the fitting regions. A limited number of transitions, marked as "Unknown" in the table, did not show a corresponding Lyman $\beta$ or $\gamma$ in the fitting regions but were required for convergence of the fit, and are thus listed here. The column densities and $b_{\rm tot}$ parameters listed in the table are obtained under the assumption that these transitions are caused by \ion{H}{I}, according to the VPFIT default.}
    \label{tab:UnrelatedInterlopers} \\
       \toprule
       Transition  & $\log({\rm N})$ & $z$ & $b_{\rm tot}$ [\kms{}] &  Transition  & $\log({\rm N})$ & $z$ & $b_{\rm tot}$ [\kms{}] \\
       \midrule
       \ion{H}{I}   &   13.725   &   3.33306   &   43.49   &   \ion{H}{I}   &   11.916  &  3.18239  &  10.58  \\
       \ion{H}{I}   &   14.148   &   3.33321   &   23.28   &   \ion{H}{I}   &   13.646  &  3.14762  &  17.90  \\
       \ion{H}{I}   &   12.940   &   3.33406   &   30.18   &   \ion{H}{I}   &   13.024  &  3.14857  &  29.28  \\
       Unknown      &   12.598   &   3.33535   &    5.34   &   \ion{H}{I}   &   12.610  &  3.14954  &  21.36  \\
       Unknown      &   12.597   &   3.33563   &    8.12   &   \ion{H}{I}   &   13.833  &  3.15078  &  23.66  \\
       Unknown      &   11.553   &   3.33589   &    1.76   &   \ion{H}{I}   &   13.421  &  3.15125  &  19.34  \\
       \ion{H}{I}   &   13.411   &   3.33581   &   54.25   &   \ion{H}{I}   &   13.606  &  3.37168  &  37.19  \\
       Unknown      &   12.272   &   3.33618   &    4.39   &   \ion{H}{I}   &   13.817  &  3.37277  &  23.06  \\
       Unknown      &   11.898   &   3.33648   &    4.61   &   \ion{H}{I}   &   13.606  &  3.37168  &  37.19  \\
       \ion{H}{I}   &   12.929   &   3.33695   &   21.45   &   \ion{H}{I}   &   13.817  &  3.37277  &  23.06  \\
       \ion{H}{I}   &   13.057   &   3.33811   &   36.28   &   \ion{H}{I}   &   13.079  &  3.37140  &  15.26  \\
       \ion{H}{I}   &   13.096   &   3.33814   &   40.18   &   \ion{H}{I}   &   12.646  &  3.37320  &  19.72  \\
       \ion{H}{I}   &   14.352   &   3.23251   &   21.80   &   \ion{H}{I}   &   12.692  &  3.37492  &  55.41  \\
       \ion{H}{I}   &   13.885   &   3.23384   &   41.77   &   \ion{H}{I}   &   12.299  &  3.37440  &   3.11  \\
       \ion{H}{I}   &   15.119   &   3.23556   &   32.92   &   \ion{H}{I}   &   12.732  &  3.37683  &  55.39  \\
       \ion{H}{I}   &   13.058   &   3.23673   &   18.99   &   \ion{H}{I}   &   11.766  &  3.37499  &  10.55  \\
       \ion{H}{I}   &   14.528   &   3.23770   &   29.51   &   \ion{H}{I}   &   11.977  &  3.37175  &   5.62  \\
       \ion{H}{I}   &   13.482   &   3.23923   &   33.15   &   \ion{H}{I}   &   11.729  &  3.37223  &   6.19  \\
       \ion{H}{I}   &   14.000   &   3.23067   &   37.35   &   \ion{H}{I}   &   12.366  &  3.12955  &  14.57  \\
       \ion{H}{I}   &   14.305   &   3.45857   &   24.65   &   \ion{H}{I}   &   15.049  &  3.13059  &  28.17  \\
       \ion{H}{I}   &   13.724   &   3.44979   &   20.32   &   \ion{H}{I}   &   13.106  &  3.13223  &  78.61  \\
       \ion{H}{I}   &   14.512   &   3.45031   &   43.96   &   \ion{H}{I}   &   13.229  &  3.35231  &  27.12  \\
       \ion{H}{I}   &   14.421   &   3.45322   &   88.19   &   \ion{H}{I}   &   12.866  &  3.35478  &  55.59  \\
       \ion{H}{I}   &   12.919   &   3.45699   &   43.84   &   \ion{H}{I}   &   12.777  &  3.35603  &  26.98  \\
       \ion{H}{I}   &   14.519   &   3.45987   &   30.85   &   \ion{H}{I}   &   12.410  &  3.35395  &  19.71  \\
       \ion{H}{I}   &   15.161   &   3.46056   &   28.97   &   \ion{H}{I}   &   12.602  &  3.35308  &  27.77  \\
       \ion{H}{I}   &   14.495   &   3.46143   &   17.85   &   \ion{H}{I}   &   11.914  &  3.35965  &  12.44  \\
       \ion{H}{I}   &   14.067   &   3.46228   &   23.03   &   \ion{H}{I}   &   11.645  &  3.35704  &  12.59  \\
       \ion{H}{I}   &   14.682   &   3.46417   &   20.98   &   \ion{H}{I}   &   13.759  &  3.11792  &  26.21  \\
       \ion{H}{I}   &   13.028   &   3.46802   &   31.45   &   \ion{H}{I}   &   13.026  &  3.11775  &  56.69  \\
       \ion{H}{I}   &   12.746   &   3.46878   &   23.68   &   \ion{H}{I}   &   12.681  &  3.12132  &  27.41  \\
       \ion{H}{I}   &   13.187   &   3.46941   &    6.16   &   \ion{H}{I}   &   12.512  &  3.11782  &  11.40  \\
       \ion{H}{I}   &   13.391   &   3.46942   &   25.08   &   \ion{H}{I}   &   12.566  &  3.11951  &  20.98  \\
       \ion{H}{I}   &   13.907   &   3.46440   &   41.01   &   \ion{H}{I}   &   12.133  &  3.12037  &  15.36  \\
       \ion{H}{I}   &   12.853   &   3.45565   &   29.48   &   \ion{H}{I}   &   12.933  &  3.11858  &  32.72  \\
       \ion{H}{I}   &   14.793   &   3.45332   &   36.89   &   \ion{H}{I}   &   12.711  &  3.11682  &  36.77  \\
       \ion{H}{I}   &   13.088   &   3.46294   &   37.56   &   \ion{H}{I}   &   12.087  &  3.11560  &  21.49  \\
       \ion{H}{I}   &   12.318   &   3.46555   &   20.00   &   \ion{H}{I}   &   12.919  &  3.09982  &  90.19  \\
       \ion{H}{I}   &   11.745   &   3.45483   &    8.14   &   \ion{H}{I}   &   13.998  &  3.10029  &  23.13  \\
       \ion{H}{I}   &   13.224   &   3.45219   &   19.20   &   \ion{H}{I}   &   12.387  &  3.10095  &  14.14  \\
       \ion{H}{I}   &   13.684   &   3.17379   &   28.70   &   \ion{H}{I}   &   12.624  &  3.10114  &  16.75  \\
       \ion{H}{I}   &   13.983   &   3.17480   &   22.47   &   \ion{H}{I}   &   13.323  &  3.10202  &  37.07  \\
       \ion{H}{I}   &   14.200   &   3.17562   &   25.02   &   \ion{H}{I}   &   13.491  &  3.10322  &  24.13  \\
       \ion{H}{I}   &   13.688   &   3.17643   &   28.67   &   \ion{H}{I}   &   12.267  &  3.10436  &  19.62  \\
       \ion{H}{I}   &   13.295   &   3.17753   &   15.13   &   \ion{H}{I}   &   12.370  &  3.10340  &  11.60  \\
       \ion{H}{I}   &   14.373   &   3.17777   &   40.62   &   \ion{H}{I}   &   11.603  &  3.10194  &   4.46  \\
       \ion{H}{I}   &   13.669   &   3.17913   &   23.71   &   \ion{H}{I}   &   12.971  &  3.10031  &  10.64  \\
       \ion{H}{I}   &   13.726   &   3.18095   &   33.59   &   \ion{H}{I}   &   12.213  &  3.10056  &   4.62  \\
       \ion{H}{I}   &   13.075   &   3.18197   &   19.54   &   \ion{H}{I}   &   12.981  &  3.32722  &  26.49  \\
       \ion{H}{I}   &   13.458   &   3.18294   &   28.16   &   \ion{H}{I}   &   13.716  &  3.32823  &  24.11  \\
       \ion{H}{I}   &   12.910   &   3.18366   &   32.69   &   \ion{H}{I}   &   13.051  &  3.33003  &  31.69  \\
       \ion{H}{I}   &   13.479   &   3.18502   &   22.77   &   \ion{H}{I}   &   13.107  &  3.33111  &  35.49  \\
       \ion{H}{I}   &   12.429   &   3.17366   &   10.30   &   \ion{H}{I}   &   13.143  &  3.32896  &  29.62  \\
       \ion{H}{I}   &   12.178   &   3.18309   &    8.68   &   \ion{H}{I}   &   13.082  &  3.32730  &  24.75  \\
       \ion{H}{I}   &   12.991   &   3.17826   &   14.57   &                &           &           &         \\
       \bottomrule
\end{longtable}